%% file: paper.tex
\documentclass[12pt,preprint]{aastex}
\usepackage{emulateapj5}
\usepackage{graphics,graphicx}
\usepackage{fancyheadings}
\usepackage{color}
\usepackage{natbib}

\newcommand{\chandra}{\textit{Chandra}}
\def\arcsec{\hbox{$^{\prime\prime}$}}
\def\deg{\hbox{$^\circ$}}

\shorttitle{Radio Galaxy 4C+29.30 in X-rays}
\shortauthors{Siemiginowska et al.}

\begin{document}

\title{Deep \chandra\ X-ray Imaging of a Nearby Radio Galaxy 4C+29.30: X-ray/Radio Connection}

\author{Aneta Siemiginowska$^{1}$, \L ukasz Stawarz$^{2,\,3}$, Chi~C. Cheung$^{4}$, Thomas L. Aldcroft$^1$, Jill Bechtold$^5$, D.~J. Burke$^1$, Daniel Evans$^{1,\,6}$, Joanna Holt$^7$, Marek Jamrozy$^{3}$, Giulia Migliori$^1$}

\affil{$^1$ Harvard Smithsonian Center for Astrophysics, 60 Garden St, Cambridge, MA 02138, USA}
\affil{$^2$ Institute of Space and Astronautical Science, JAXA, 3-1-1 Yoshinodai, Sagamihara, Kanagawa, 229-8510, Japan} 
\affil{$^3$ Astronomical Observatory, Jagiellonian University, ul. Orla 171, 30-244 Krak\'ow, Poland}
\affil{$^4$ National Research Council Research Associate, National Academy of Sciences, Washington, DC 20001, resident at Naval Research Laboratory, Washington, DC 20375, USA}
\affil{$^5$ Steward Observatory, University of Arizona, 933 North Cherry Avenue, Tucson, AZ 85721, USA}
\affil{$^6$ Elon University, Department of Physics, 100 Campus Drive, Elon, NC 27244, USA}
\affil{$^7$ Leiden Observatory, Leiden University, PO Box 9513, 2300 RA Leiden, The Netherlands}

\smallskip
\email{asiemiginowska@cfa.harvard.edu}


\label{firstpage}

\begin{abstract}

  We report results from our deep \chandra\ X-ray observations of a
  nearby radio galaxy, 4C+29.30 ($z=0.0647$). The \chandra\ image
  resolves structures on sub-arcsec to arcsec scales, revealing
  complex X-ray morphology and detecting the main radio features: the
  nucleus, a jet, hotspots, and lobes.  The nucleus is absorbed
  ($N_{\rm H} \simeq 3.95^{+0.27}_{-0.33} \times 10^{23}$\,cm$^{-2}$)
  with an unabsorbed luminosity of $L_{\rm 2-10\,keV} \simeq (5.08 \pm
  0.52) \times 10^{43}$\,erg\,s$^{-1}$ characteristic of Type 2 AGN.
  Regions of soft ($<2\,\rm keV$) X-ray emission that trace the hot
  interstellar medium (ISM) are correlated with radio structures along
  the main radio axis indicating a strong relation between the two.
  The X-ray emission beyond the radio source correlates with the
  morphology of optical line-emitting regions.  We measured the ISM
  temperature in several regions across the galaxy to be $kT \simeq
  0.5$\,keV, with slightly higher temperatures (of a few keV) in the
  center and in the vicinity of the radio hotspots.  Assuming these
  regions were heated by weak shocks driven by the expanding radio
  source, we estimated the corresponding Mach number of 1.6 in the
  southern regions. The thermal pressure of the X-ray emitting gas in
  the outermost regions suggest the hot ISM is slightly
  under-pressured with respect to the cold optical-line emitting gas
  and radio-emitting plasma, which both seem to be in a rough pressure
  equilibrium. We conclude that 4C+29.30 displays a complex view of
  interactions between the jet-driven radio outflow and host galaxy
  environment, signaling feedback processes closely associated with
  the central active nucleus.

\end{abstract}


\section{Introduction}
\label{sec:intro}

Recent observational and numerical studies highlight the importance of
active galactic nuclei (AGN) induced feedback on the co-evolution of a
super-massive black hole (SMBH) and its host galaxy
\citep{silk1998,merloni2008}.  Various physical processes involved are
in many respects only poorly understood, although the emerging
agreement is that this feedback must have played a significant role
during the primary epochs of galaxy formation at high redshifts
\citep{croton2006}.

The impact of a SMBH on the larger scale environment has been
evidenced by powerful AGN outflows \citep[see e.g.,][]{tadhunter2007}.
In particular, it has been established that substantial amounts of
energy have been deposited to the intergalactic medium by powerful
radio-emitting outflows associated with massive elliptical galaxies
located in galaxy clusters \citep[e.g.,][]{nulsen2005}, and that
relatively dense optical line-emitting gas participates in galaxy
scale outflows \citep[e.g.,][]{morganti2005,holt2008}.  The highest
angular resolution X-ray observations performed by the \chandra\ X-ray
Observatory (\chandra) proved to be particularly important in studying
the physics of the jet-related feedback, since they provided
direct evidence for the interactions between the radio-emitting
outflows and surrounding gas, most prominently in the case of
clusters of galaxies \citep{mcnamara2007}, but also in central regions
of nearby galaxies \citep{hardcastle2007,wangfabiano,wang2011}.
However, only in a few cases could the details of such interactions be
accessed, with a complex picture involving recurrent AGN outbursts shaping
the properties of the interstellar and intergalactic medium at
different evolutionary stages of the systems being revealed.

Among a few sources where the jet-ambient medium interactions 
(particularly within the confines of the host galaxies) can be
traced with sufficient details at X-ray frequencies are the well-known radio
galaxies M~87 and NGC~1275
\citep[e.g.,][respectively]{million2010,fabian2011} and the brightest
nearby Seyfert galaxy NGC\,4151 \citep{wang2010}. The nearest radio-loud AGN,
Centaurus~A (distance $D \simeq 3.7$\,Mpc) is a well-known example 
where the rejuvenated radio structure and jet-ambient medium
interactions can easily be mapped at various scales and different
frequencies \citep[e.g.,][and references therein]{morganti2010}. In
all these cases, the X-ray observations exposed signatures of shocks or
shock heated medium, ionization of the gas, the scales of the impact of
the radio source on the medium and most importantly, the overall
energetics of the systems.  Such investigations of the feedback
processes at various stages of the AGN activity are highly desirable,
however, they can only be performed for nearby galaxies given the
capabilities of the currently operating X-ray missions.

To investigate such feedback processes, we selected the low-redshift
radio galaxy 4C+29.30 (0836+299, $z=0.0647$, $D \simeq
287$\,Mpc)\footnote{We assume the following cosmological parameters:
  $H_{\rm 0} = 71$\,km\,s$^{-1}$\,Mpc$^{-1}$, $\Omega_{\rm M} = 0.27$,
  $\Omega_{\rm vac} = 0.73$.} for an in-depth X-ray study with
\chandra\,.  At optical frequencies, the moderate-luminosity of its
nucleus resembles that of Seyfert type 2 galaxies.  The radio source
displays a transitional FR\,I/FR\,II radio morphology and intermediate
radio luminosity of the order of $\sim 10^{42}$\,erg\,s$^{-1}$. It is
hosted by an elliptical galaxy characterized by a distorted structure
due to a relatively fresh merger with a gas-rich companion
\citep{vanbreugel1986}.  The radio structure extends out to $\sim
30$\,kpc from the center of the galaxy and consists of a prominent
one-sided jet and lobe visible to the south, and a counter-lobe to the
north. No counterjet has been detected at radio frequencies indicating
an intermediate jet inclination to the line of sight and relativistic
beaming involved.  4C+29.30 is particularly interesting due to its
prominent ($\sim 10^{43}$\,erg\,s$^{-1}$) and extended ($\sim
30$\,kpc) emission-line regions intertwined with the described radio
structure on scales that can be resolved with \chandra\ (1$\arcsec =
1.227$\,kpc at the redshift of the source). As argued in
\citet{vanbreugel1986},  this complex morphology and the overall
  energetics of the line-emitting gas suggest strong interactions
  between the radio outflow and the cold clumpy phase of the gaseous
  environment in this system, which is therefore specially well-suited
  for the observational investigation of the feedback process.

Similarly to the case of Centaurus~A, the inner radio structure and
the host galaxy of 4C+29.30 are surrounded by a $\sim 600$\,kpc
diffuse radio emission which extends in the N-S direction
\citep{jamrozy2007}.  The primary (inner/young) and outer radio
structures are relatively well-aligned, as observed in the class of
so-called `double-double' radio galaxies \citep[see][for a review of sources of
recurrent jet activity in general]{saikia2009}. The presence of these outer lobes
indicate a highly intermittent jet activity in the source with the
young ($\sim 33$\,Myr) kpc-scale outflow propagating within relic
lobes formed in a previous epoch of the jet activity (some $\gtrsim
200$\,Myrs ago).  We note that modulated jet activity resulting
in the formation of multiple radio lobes was also claimed in the cases
of the radio galaxies M~87 and NGC~1275 located in rich clusters
\citep[e.g.,][]{owen2000,fabian2000}, albeit on scales much smaller than
that of the outer radio halos in 4C+29.30 and Centaurus~A.

In the X-rays, 4C+29.30 was not detected by ROSAT \citep{canosa1999}
and the first such detection of the source was obtained by \chandra.
This initial \chandra\ observation from April 2001 (ObsID=2135) was
short ($\sim 8$\,ksec), but provided a snap-shot view of the source's
complex X-ray morphology, including the X-ray detection of the nucleus
and radio hotspots, and possible indication of an X-ray jet and
diffuse components as well
\citep{gambill2003,sambruna2004,jamrozy2007}.  However, this early
observation did not provide constraints on the energetics of the
system or the nature of the X-ray emission in different regions due to
the very limited photon statistics.

To address the above issues and to discuss in more detail the
jet-related feedback in 4C+29.30, we obtained new deep high angular
resolution X-ray observations of the source with \chandra.  The new
\chandra\ data provide the high quality X-ray imaging of a radio
source with an extended optical emission-line region revealing their
complex relationship. The results of our study are presented here.  In
the following, we begin by discussing the technical settings of the
observation and analysis methods in Section~\ref{sec:chandra}. We then
show details of the X-ray morphology in Section~\ref{sec:x-morphology}
and the results of spectral analysis of individual X-ray features in
Section~\ref{sec:spectra}. We finally discuss and summarize the main
results in Sections~\ref{sec:discussion} and ~\ref{sec:summary},
respectively.

\section{Chandra Observations and Analysis Methods}
\label{sec:chandra}

Deep \chandra\ ACIS imaging observations of 4C+29.30 were performed in
February 2010.  The approved 300\,ksec observation was split into four
separate pointings that sum to a total exposure of 286.4\,ksec (see
Table~\ref{table-1} for details).  The source was placed on the
back-illuminated CCD (S3) and was offset by $-1$\,arcmin in Y
coordinates to ensure that the diffuse galaxy emission was
positioned away from a chip gap. The observations were made in VFAINT mode and
the full-window mode.  The background was quiet and the observations
were not affected by Solar flares.  The four pointings were performed
with similar enough configurations that they could be merged together
for the purpose of image analysis. Note that the exposure maps are
flat on the scale size of the galaxy in all four observations.
Analysis was performed with the CIAO version 4.3 software
\citep{ciao2006} using CALDB version 4.4. All spectral modeling was
done in Sherpa \citep{freeman2001,refsdal2009}.  We used the Cash and
Cstat fitting statistics \citep{cash1979} and the Nelder-Mead
optimization method \citep{neldermead1965}.

We processed the data by running CIAO tool {\tt acis-process-events}
and applied the newest calibration files, filtered VFAINT background
events, and ran a sub-pixel event-repositioning algorithm (and set {\tt
  pix\_adj=EDSER}). This final step provides the highest angular
resolution X-ray image data for the most up-to-date ACIS-S
calibration.  While the four separate observations were merged into one
in order to obtain the best quality image for studying the X-ray
morphology, for all spectral modeling, we extracted the spectra and
response files from the individual observations.

\section{X-ray Morphology}
\label{sec:x-morphology}

\begin{figure*}
\begin{center}
\includegraphics[width=6.in]{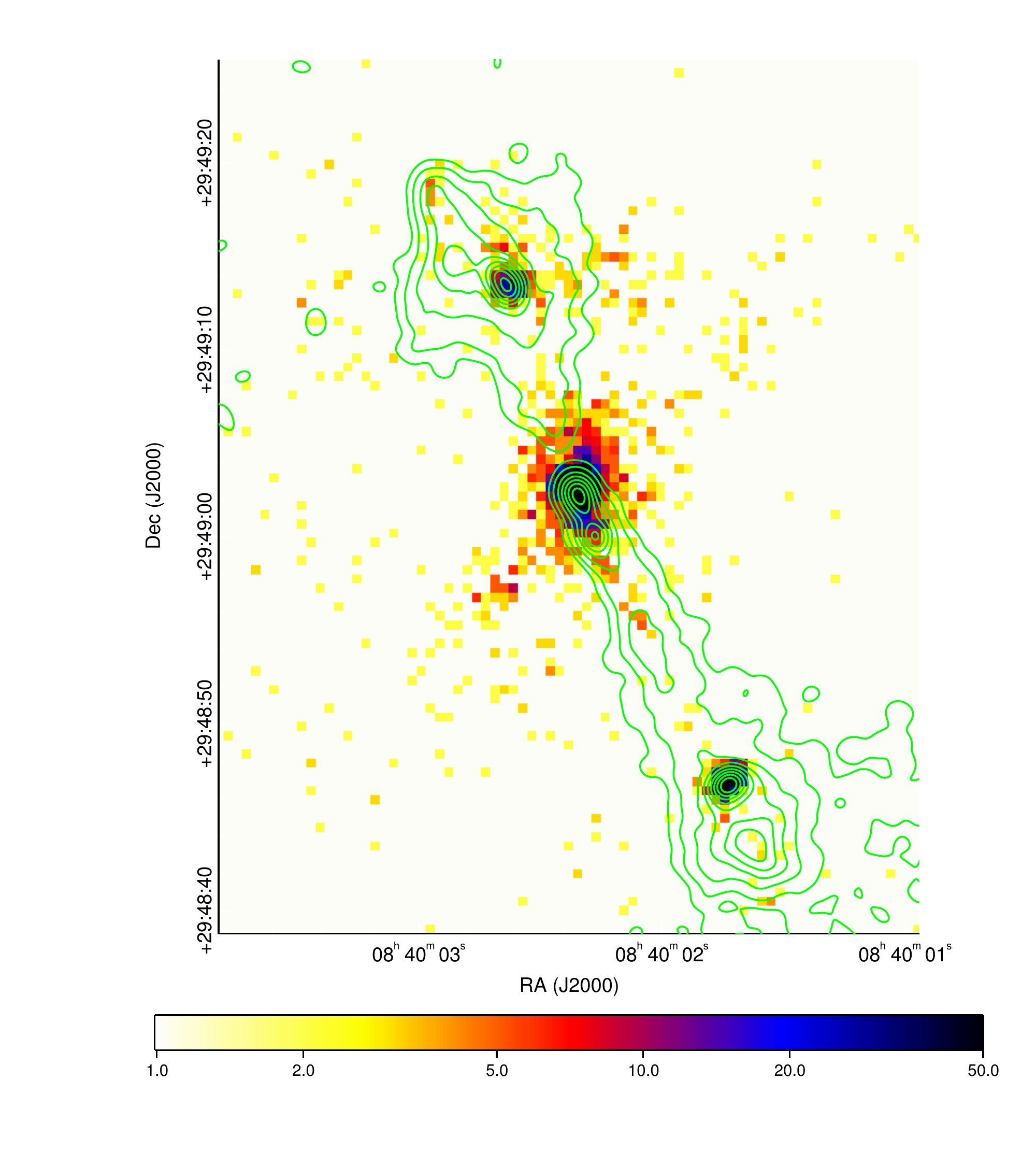}
\caption{\small \chandra\ ACIS-S image of the 4C+29.30 radio galaxy in
  the $0.5-7$~keV energy band.  The four individual observations have
  been merged into one image which was binned to show the scale of
  ACIS-S pixels, 1\,pixel$ = 0.492\arcsec$. The image is overlayed
  with the contours from VLA 1.45\,GHz map of \cite{vanbreugel1986}
  obtained from the NRAO archives and processed by \cite{cheung2004}. The
  contours are the fractions of the highest contour (0.04, 0.042,
  0.044, 0.12, 0.24, 0.48)$\times$0.005\,Jy/beam with a clean beam
  size of $1\arcsec \times 1$\arcsec.  The color bar indicates
  a number of counts per pixel.
}
{\label{acis}}
\end{center}
\end{figure*}

\begin{figure*}
\begin{center}
  \includegraphics[width=5.5in]{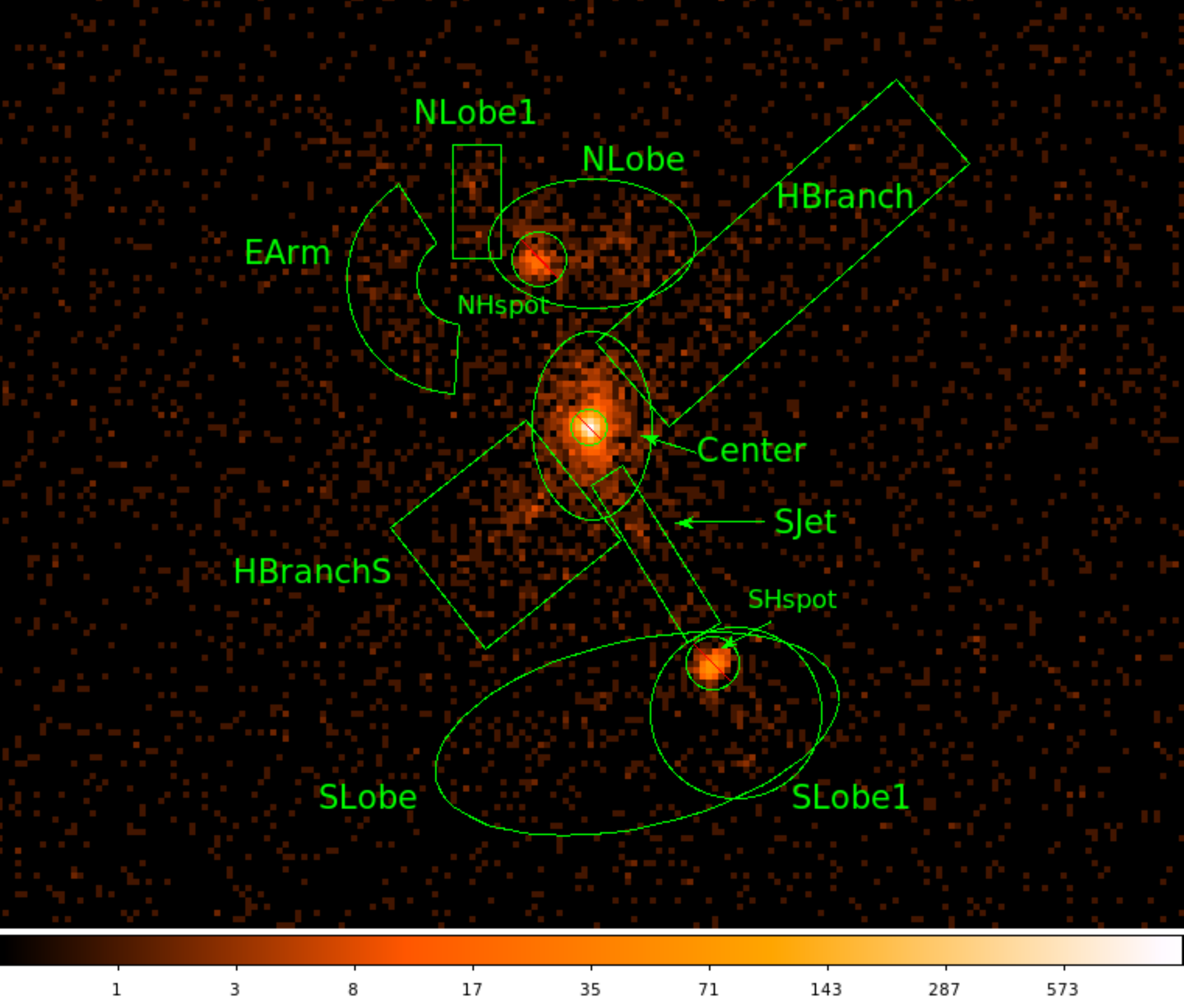}
  \caption{\small {\it Chandra\/} ACIS-S image of the 4C+29.30 radio
    galaxy in the $0.5-7$ keV energy range. Components of the emission
    are marked by names and regions.  The same regions were used for
    spectral extractions. (Note that some regions were excluded in
    extraction of the diffuse emission and they have been marked by
    red lines.)  The pixel size is equal to 0.492\arcsec, i.e., the
    ACIS pixel size.}  {\label{spec-regions}}
\end{center}
\end{figure*}

Figure~\ref{acis} shows our new \chandra\ ACIS-S X-ray image of the
galaxy 4C+29.30 obtained from the summed $286.4$\,ksec exposure. The
image is overlayed with the radio contours from VLA 1.45~GHz map of
\cite{vanbreugel1986}.  A complex morphology with many regions of
enhanced emission is displayed in this image.  The X-ray emission
spans about $\sim 60$\,kpc in total extent and is marked by a
pronounced center and two bright components coincident with the jet
termination regions (the radio `hotspots'). The low surface brightness
diffuse X-ray emission connecting the hotspots along the jet axis,
together with the previously unknown linear X-ray feature almost
perpendicular to the radio axis (in the NE-SW direction), creates an
overall impression of an X-like shape for the diffuse X-rays.  We
identify and mark these various X-ray emission regions in
Figure~\ref{spec-regions}.

\begin{figure*}
\begin{center}
\includegraphics[width=5.5in]{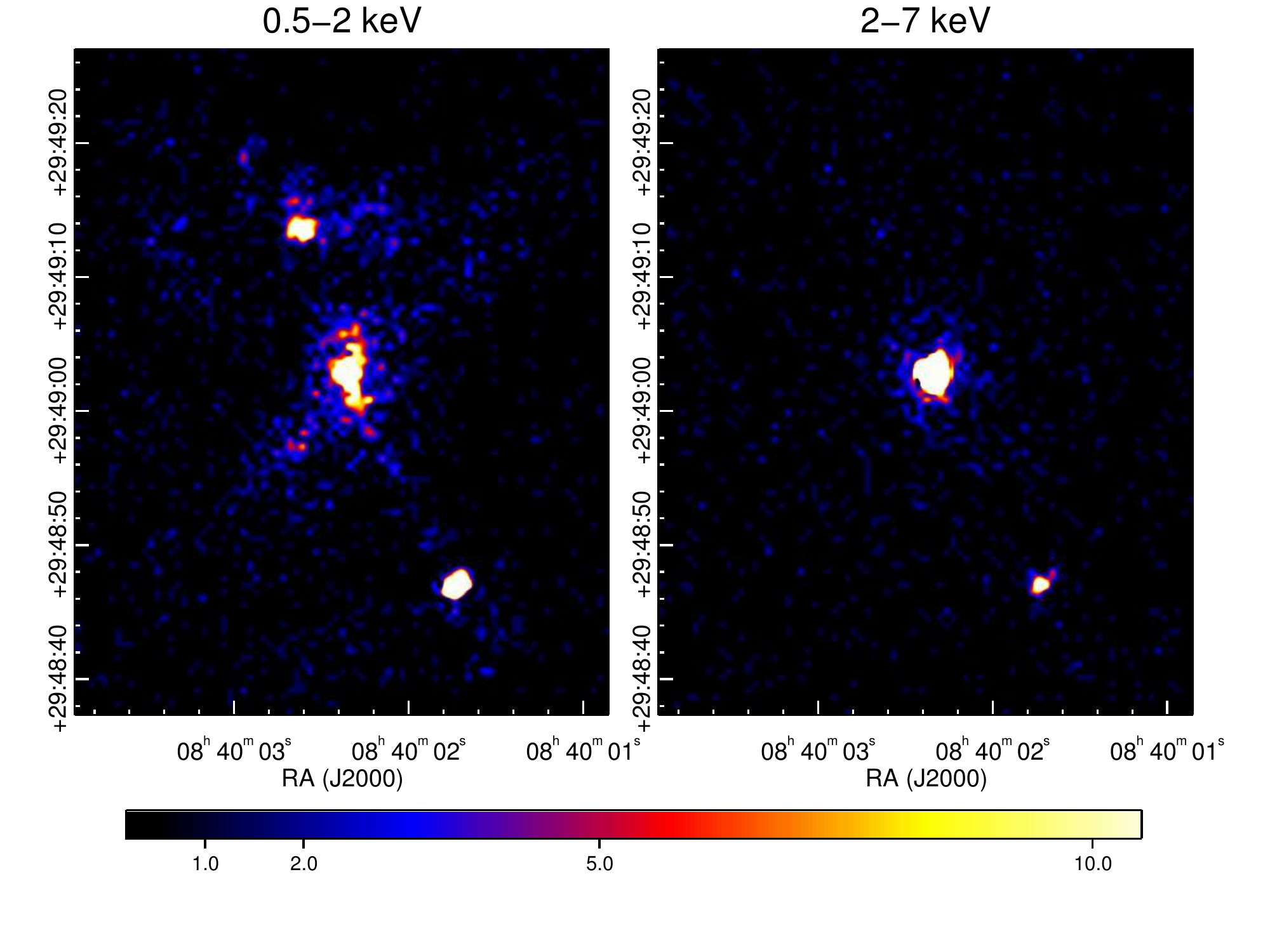}
\caption{\small {\it Chandra\/} ACIS-S images of the 4C+29.30 radio
  galaxy in the soft ($0.5-2$~keV; left) and hard ($2-7$~keV; right)
  band. The images were smoothed with the Gaussian kernel with
  $\sigma=0.75\arcsec$. The color scale indicates number of counts as
  marked in the color bar.  The images are on the same physical scale.
  The core and the southern hotspot are prominent in the hard band
  image.}  {\label{soft}}
\end{center}
\end{figure*}

Figure~\ref{soft} presents the smoothed X-ray images in the soft
($0.5-2$\,keV) and hard ($2-7$\,keV) bands.  As seen in the images,
most of the diffuse structure is suppressed above $\sim 2$\,keV. The
dominant core (hereafter denoted as `Core') and the Southern hotspot
(`SHspot') are in fact the only features detected at photon energies
$>2$\,keV (note however the dramatically decreased prominence of
SHspot in the hard band).  In contrast, the soft band image shows a
complex and rich morphology with several distinct features, either
relatively compact blobs or extended filaments, and the quite
prominent horizontal branch perpendicular to the main axis of the
radio source.

\begin{figure*}
\begin{center}
\includegraphics[width=6.5in]{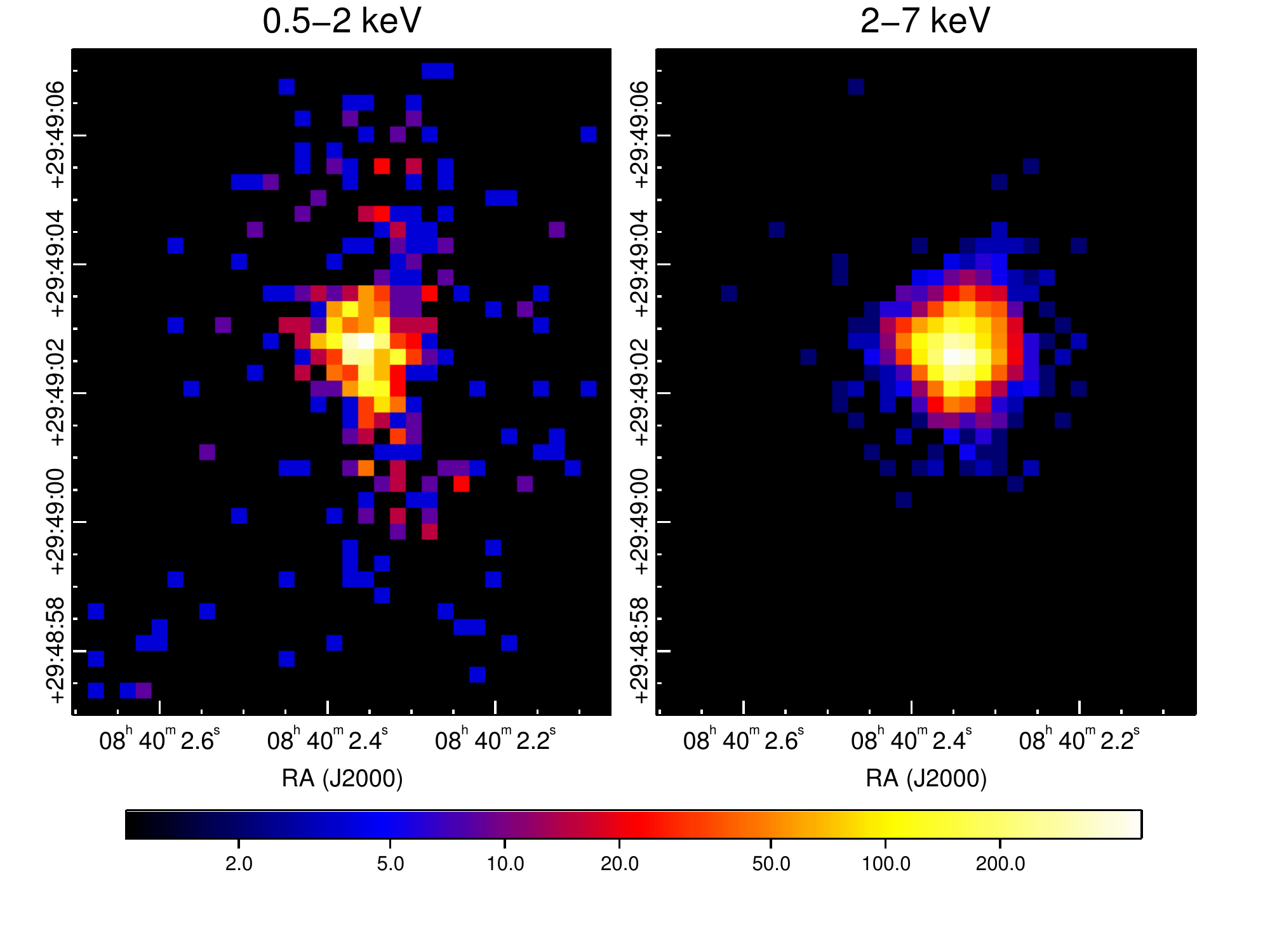}
\caption{\small ACIS-S images of the core regions in two bands, soft
  $0.5-2$~keV and hard $2-7$~keV. The pixel size is 0.242\,arcsec 
(0.5 of the original ACIS-S pixel size). The color scale indicates a
 number of counts in the image pixel. 
}
 {\label{core-narrow-band}}
\end{center}
\end{figure*}

Figure~\ref{core-narrow-band} shows the soft and hard X-ray images of
the central region of the galaxy (`Center').  The soft emission is
diffuse within the inner 8\arcsec\.  The hard emission is concentrated
within a circular region with 1.5\arcsec\ radius centered on the
nucleus, with no sign of any extended emission component, thus is most
likely dominated by the central AGN.

\begin{figure*}
\begin{center}
\includegraphics[width=\columnwidth]{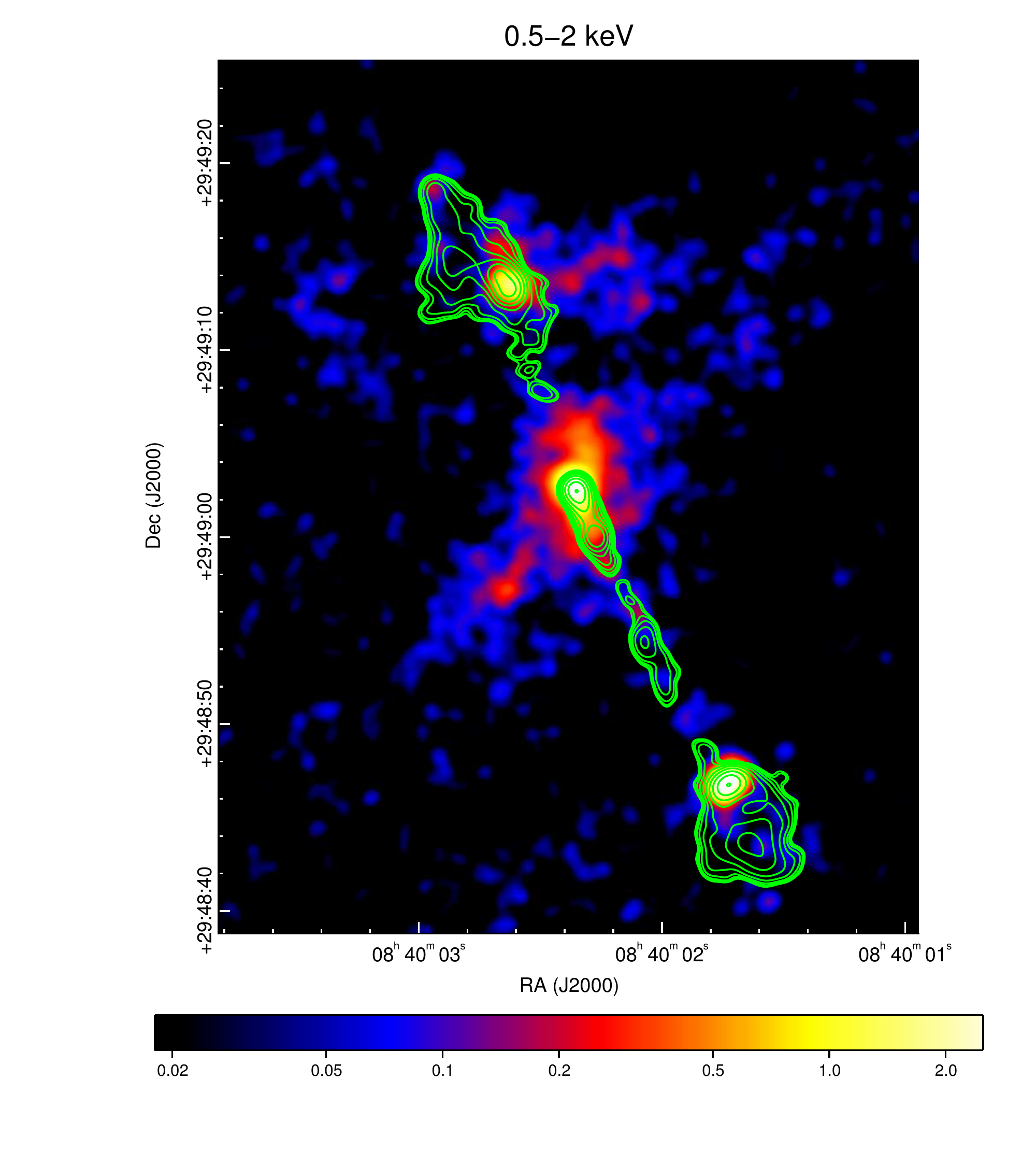}
\caption{\small Smoothed \chandra\ ACIS-S image overlaid with contours
  from VLA map at 5~GHz. Only X-ray photons from $0.5-2$\,keV range
  were included in the image. The original image was binned by a 1/4 of
  the original ACIS-S pixel size and then smoothed with the Gaussian function
  ($\sigma=0.5\arcsec$). The utilized radio map was initially
  presented in \cite{sambruna2004} using archival data originally
  published by \cite{vanbreugel1986}.  The scale is indicated in the
  bottom color bar. 
}{\label{xradio}}
\end{center}
\end{figure*}

In Figure~\ref{xradio} we show radio contours from a VLA 5\,GHz image
overlaid on the smoothed $0.5-2$\,keV ACIS-S image.  The linear X-ray
emission component extending from the core to the south (hereafter
`SJet') is closely aligned with the radio jet.  Note that the SJet
feature covers a much broader area than the well-defined radio jet,
and hence it is not obvious if all the X-rays are non-thermal in
origin. The radio jet terminates at a hotspot $\sim 20\arcsec$ away
from the core, where the whole (radio) outflow suddenly decollimates
and extends further to the south in a much less organized fan-like
pattern (see also Figure~\ref{acis}).  The X-ray emission aligned with
the radio jet flares similarly at the position of SHspot, but it is
not clear if the X-rays follow the radio morphology also further to
the South (`SLobe1' region).

\begin{figure*}
\begin{center}
\includegraphics[width=5in]{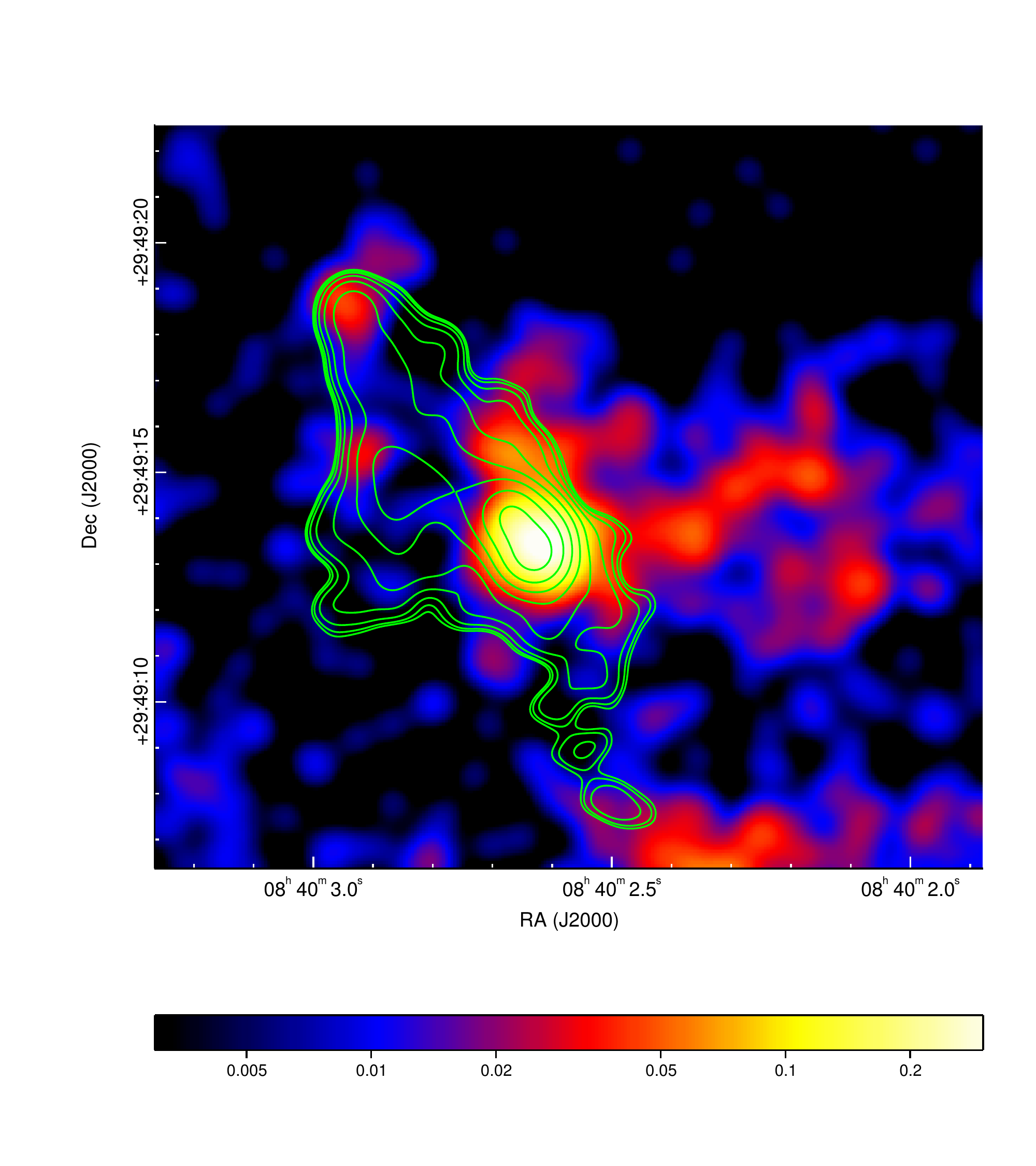}
\caption{\small ACIS-S image of the northern lobe and hotspot
  regions in the energy range $0.5-2$~keV overlaid with radio contours from
  the VLA 5~GHz image. The X-ray image was binned to 0.125\arcsec\ pixel size and
  smoothed with the Gaussian with $\sigma=0.5\arcsec$.}
  {\label{lobe-narrow-band}}
\end{center}
\end{figure*}

The X-ray emission component detected to the north of the core extends
over a 30\arcsec\ wide band roughly perpendicular to the main radio
source axis.  This emission has a rather complex morphology and
coincides roughly with the position of an extended optical
emission-line region analyzed in detail by \citet{vanbreugel1986}.
Here, we can identify the X-ray counterpart of the northern radio
hotspot (hereafter `NHspot'), as well as its extension to the west
forming a lower-surface brightness tail (`NLobe').  This tail may
however simply be a superposition of some smaller-scale enhancements
visible in the field rather than a truly coherent feature. Indeed, a
zoomed-in view of the northern region presented in
Figure~\ref{lobe-narrow-band} shows that the X-ray emission beyond
the NHspot radio extent forms an apparent wiggly filamentary feature
approximately perpendicular to the radio axis.

In general, it appears that the X-ray emission in the northern region
coincides better with the line-emitting gas than with the
radio-emitting plasma, and that the X-ray emission even ``hugs'' the
northern radio lobe on its eastern side from the outside (`EArm'
component in Figure~\ref{spec-regions}). It is therefore possible
that the bulk of the X-ray emission here could be associated
with the thermal gas that is being shocked by the expanding radio
source, rather than with the non-thermal emission of relativistic
plasma injected by the radio jet into the expanding lobes. This idea is
strongly supported by the fact that the brightest northern X-ray
features (with the exception of NHspot) are located near the
boundaries of the radio structures, and also by the spectral analysis
presented below (Section~\ref{sec:spectra}). 

We identified two X-ray features `HBranchS' and 'HBranch' located
almost perpendicular to the radio source axis. There is no obvious
correspondence between these features and the line-emitting gas. A
faint optical filament of a similar (projected) size as the radio jet,
but misaligned (by about $\sim20\deg$) with the jet axis
\citep{vanbreugel1986} is oriented in between the jet and the
`HBranchS'. However, prominent dust lanes seen in the optical HST
image are located in the similar direction to these X-ray features
suggesting a possible relation between the two. The optical studies
support an idea of a merger event in which a smaller rich galaxy was
captured and the dust shells are the remnants from that event.  Thus
the observed X-ray morphology appears to be shaped by a complex
mixture of a merger and a radio source activity, and both influencing
the properties of the ISM.

\section{X-ray Spectral Properties}
\label{sec:spectra}

We extracted X-ray spectra and created calibration response files
for each of the four individual \chandra\ observations using the
specified regions marked on Figure~\ref{spec-regions} and listed in
Table~\ref{table-2}. The response files are based on CALDB 4.4 and
include the updated ACIS-S contamination
model\footnote{\texttt{http://cxc.harvard.edu/ciao/why/acisqecontam.html}}.  
The spectra from the four observations were fitted simultaneously in Sherpa
using Cash statistics with the {\tt simplex-neldermead} and {\tt
  moncar} methods. The background and source spectra were fit 
simultaneously assuming several different parametrized models (see 
below). We used the blank-sky background data to create a background 
model that was then scaled to our observations \citep[for details, 
see][]{siemi2010}.  For the features embedded in the diffuse emission, 
we also included background model components that accounted for the 
diffuse emission contributing to the source spectrum.  The modeling 
details for all the features are described below.

We modeled the spectra of all the features defined in
Section~\ref{sec:x-morphology} assuming two main emission models: an
absorbed power-law model and an absorbed thermal emission model.  In
the latter case we used the APEC\footnote{Model of the emission from
  collisionally ionized plasma based on ATOMDB,
  \texttt{http://www.atomdb.org/}} thermal model describing emission
from collisionally ionized plasma, which allows for varying metal
abundances. For most of the regions fitted with this model we found
low values of the abundance parameter, and for the regions located
further away from the galaxy center we could only determine the upper
limits ($< 10\%$ solar). Similar abundance profiles are often found in
elliptical galaxies. On the other hand there are many systematic
issues related to fitting the abundances in multi-phase ISM using low
signal to noise data \cite[see][for details]{dongwoo2012}.  For this
reason we favor the best fit model parameters emerging from the
thermal fits consisting of a pure thermal bremsstrahlung emission,
obtained by setting the abundance parameter in the APEC model to 0.
This choice reduces the number of free model parameters critical in
modeling limited photon statistics data. However, for completeness, we
provide the best fit APEC model parameter values obtained with the
abundance parameter set free to vary. The best-fit parameters for all
the regions, including the resulting unabsorbed X-ray fluxes in the
soft ($0.5-2$\,keV) and hard ($2-10$\,keV) bands, for both models are
summarized in Tables~\ref{table-4} and \ref{table-5}, respectively.
The model results for the nucleus, which required a more sophisticated
treatment, are given separately in Table~\ref{table-3}. The quoted
errors are 1$\sigma$ for one significant parameters, unless
specifically noted.

\subsection{The Nucleus}

Based on the soft and hard X-ray band images of the source, it is 
apparent that the unresolved nucleus, unlike the diffuse structure 
surrounding it, is bright at photon energies exceeding about 2\,keV.
Utilizing the hard band image ($2-7$\,keV; see 
Figure~\ref{soft}), we measured a centroid for the nucleus at 
$\alpha$(J2000)=$08^h40^m02^s.3$, $\delta$(J2000)=$+29^o49'02''.57$. 
This position agrees with that of the galaxy centroid quoted in NED 
based on the Sloan Digital Sky Survey 
DR6~\footnote{\texttt{http://www.sdss.org/dr6/}}.

We defined the nucleus (core) region as a circle with a radius equal
to 1.25\arcsec\ centered on the above position and extracted the X-ray
spectra and response files for that region. A total of 5328$\pm73$
counts (5303.7$\pm73.2$ net counts) with photon energies between
0.5\,keV and 7.0\,keV were found. Approximately $\sim90\%$ of the
detected counts have energies between 2\,keV and 7\,keV.  We note
  that the expected pileup fraction for the counts in the nucleus is
  low, $<2\%$, and we neglected the pileup effects in our analysis.
To account for the local background contribution to the core emission,
we assumed an annulus region centered on the core position with an
inner radius of 1.5\arcsec\ and an outer radius of 10\arcsec. In
fitting the background spectrum, we assumed a model composed of a
thermal bremsstrahlung emission and power-law component, and the
resultant fitted parameters are summarized in Table~\ref{table-3}.

The \chandra\ spectrum of the core was poorly fitted with an absorbed
power law model, with $\chi^2_{\nu}=2.56$ in the standard $\chi^2$
test applied to the binned data and $\nu$=453 degrees of freedom.
Therefore, we attempted to fit the core spectrum with several
parametric models that consisted of absorbed and unabsorbed emission
components.  All models included the Galactic absorption column
density of $N_{\rm H} (\rm Gal) = 3.98\times 10^{20}$\,cm$^{-2}$
\citep{dickey1990}.  Our selection of models and their best-fit
parameters are listed in Table~\ref{table-3}. We found that the
nucleus is significantly absorbed with a relatively high intrinsic
absorption column density of $N_{\rm H} (\rm int)
=3.95^{+0.27}_{-0.33} \times 10^{23}$\,cm$^{-2}$ and that the
unabsorbed lower energy X-ray spectrum constitutes a separate emission
component. This type of a composite spectrum consisting of an
unabsorbed soft X-ray emission and a highly absorbed hard X-ray
component is often found in the centers of radio galaxies
\citep{hardcastle2009}.

To fit the hard band X-ray emission detected from the core, we assumed 
an absorbed power-law model.  In this model, the best-fit photon index 
was $\Gamma=1.70^{+0.38}_{-0.36}$
($N_{\rm H} (\rm int)$ as given above) and the resultant unabsorbed
flux of $(5.1 \pm 0.5) \times 10^{-12}$\,erg\,cm$^{-2}$\,s$^{-1}$
corresponds to a $2-10$\,keV luminosity of $(5.0 \pm 0.5) \times
10^{43}$\,erg\,s$^{-1}$.
We note that the derived values of the photon index and
luminosity of the nucleus in 4C+29.30 are similar to the ones found in the
nuclei of Seyfert galaxies \citep{singh2011}. In Section~\ref{sec:discussion} 
we discuss implications of the luminous nucleus on
the physical state of the environment in this object.

The observed soft (0.5-2\,keV) unabsorbed emission component contains
only about $\sim 10\%$ of the total counts detected from the core.  We
first tried to model this emission with an unabsorbed power-law
component which was added to the absorbed power-law model discussed
earlier, fitting both components simultaneously.  We obtained a
reasonable photon index of $\Gamma_{\rm s} = 1.60 \pm 0.12$ and a
$0.5-2$\,keV flux of $6.4 \times 10^{-15}$\,erg\,cm$^{-2}$\,s$^{-1}$
for this model. Next, instead of a power law model, we tried two types
of thermal models added to the intrinsically absorbed power-law
component: a bremsstrahlung emission and the APEC plasma model.  Both
thermal models gave a good description of the soft emission. The
best-fit temperature of the bremsstrahlung emission is equal to
$kT_{\rm b}=6.8^{+9.9}_{-2.5}$\,keV, while the APEC gave a slightly
higher, but consistent (within the errors) temperature of $kT_{\rm
  a}=9.6^{+10.0}_{-4.8}$\,keV and 3$\sigma$ upper limit to the metal
abundances of $A<1.35$ with respect to the Solar value. For the
bremsstrahlung model, we measure a $0.5-2$\,keV flux of $7.1 \times
10^{-15}$\,erg\,cm$^{-2}$\,s$^{-1}$ corresponding to a luminosity of
$6.9 \times 10^{40}$\,erg\,s$^{-1}$.  Although we were unable to
discriminate statistically between any of the above models for the
soft component, we discuss a possible origin of this soft emission in
Section~\ref{sec:discussion}.

\subsection{X-ray Jet}

We selected a narrow box region (marked in Figure~\ref{spec-regions}
as SJet) to delineate the X-ray emission to the south of the nucleus
to study the emission that could be associated with the known radio
jet at that location.  We fitted both an absorbed power-law and an
absorbed thermal model to the X-ray spectrum of this region (see
Tables~\ref{table-4} and \ref{table-5}, respectively). In the
power-law model, we obtained a photon index, $\Gamma = 2.2\pm0.2$,
when we fixed the absorption parameter at the Galactic value, and
obtained only upper limits to both the absorbing column and photon
index when the absorption parameter was left free in the fit.  The
upper limit to the absorption is consistent with the Galactic column
thus we conclude that there is no intrinsic absorption of the X-ray
emission from this jet region.  In the framework of the thermal model,
we obtained the best fit temperature of $kT_{\rm b} =
2.1^{+0.9}_{-0.6}$\,keV ($68\%$ uncertainties for 1 significant
parameter).   We could only obtain a 3$\sigma$ upper limit of $A
  <0.13$ of the Solar values for the metal abundances in this region.

The total luminosity in the jet assuming the best-fit power-law model
is equal to 1.4$\times10^{40}$\,erg\,s$^{-1}$ in the $0.5-2$\,keV
range, which is comparable to the luminosity in the hard band
($2-10$\,keV) of $1.2 \times 10^{40}$\,erg\,s$^{-1}$.  The thermal versus
non-thermal origin for the X-ray emission detected from SJet is
discussed in Section~\ref{sec:discussion}.

\subsection{Hotspots}

There are two enhanced regions of X-ray emission in the \chandra\
image located near the opposite extremes of the radio source. They
appear to be associated with the radio hotspots, although the northern
radio feature does not display a prominent hotspot in the VLA map.
The X-ray images shown in Figure~\ref{soft} indicate that the northern
hotspot emission is rather soft, as it is very faint in the hard band
image. In contrast, the hotspot to the south is quite prominent in the
the hard band image. We extracted the spectra of both features
assuming circular regions with 1.2\arcsec\ radius each and fit the
data assuming a single absorbed power-law model and a thermal emission
model. In both cases, we both fixed the absorption column at the
Galactic value and set it free in the fits.

\subsubsection{Southern Hotspot - the end of a Jet}

The best-fit model parameters for the southern hotspot are listed in
Tables~\ref{table-4} and \ref{table-5}. The best-fit power-law photon
index for the model with Galactic absorption only
($\Gamma=2.16\pm0.08$) is consistent with the previous determination
based on the analysis of the original shorter 8\,ksec \chandra\ data
\citep{sambruna2004,jamrozy2007}, but now with improved photon
statistics. When the absorption was allowed to vary in the fit, we
obtained a value of the intrinsic column density, $N_{\rm H} =
(6.0\pm4.0) \times 10^{20}$\,cm$^{-2}$, consistent with the Galactic
column, and a photon index ($\Gamma=2.24^{+0.15}_{-0.16}$) similar to
the above value.  The resulting model fluxes in the soft and the hard
bands are equal to $F_{\rm 0.5-2 keV} = 7.75 \times
10^{-15}$\,erg\,cm$^{-2}$\,s$^{-1}$ and $F_{\rm 2-10 keV} = 6.27
\times 10^{-15}$\,erg\,cm$^{-2}$\,s$^{-1}$, corresponding to
luminosities of $7.7 \times 10^{40}$\,erg\,s$^{-1}$ and $6.2 \times
10^{40}$\,erg\,s$^{-1}$, respectively.  Besides the nucleus, the
  southern hotspot is the only feature with significant emission above
  $2$\,keV, thus signaling non-thermal processes at work, as in fact
  expected at the termination region of a relativistic jet, or a
  presence of a particularly hot X-ray emitting gas in the immediate
  hotspot vicinity.  Indeed, for the thermal model fit, the fitted
  temperature appears relatively high with $kT_{\rm a} =
  2.3^{+0.3}_{-0.2}$\,keV and a 3$\sigma$ upper limit for the metal
  abundances of $<0.09$ with respect to Solar.

Next, we modeled the X-ray emission immediately outside the southern
hotspot (region SLobe1), as well as within the more extended diffuse
emission to the south-east (SLobe; see Tables~\ref{table-4} and
\ref{table-5}). For the power-law model, the fitted X-ray spectrum of
SLobe1 is characterized by a photon index of
$\Gamma=2.09^{+0.28}_{-0.26}$, which is very similar to that of
SHspot. The best-fit thermal model for this emission component shows a
relatively high temperature of $kT_{\rm b} =
2.24^{+1.40}_{-0.74}$\,keV. We cannot discriminate between a thermal
and non-thermal origin of this emission statistically because of the
low signal-to-noise data.  However, from the physical point of view,
the emission of SLobe1 corresponding to the soft and hard luminosities
roughly $10^{40}$\,erg\,s$^{-1}$, could well be a mixture of both
thermal and non-thermal photons (see Section~\ref{sec:discussion}
below).

On the other hand, the best fit model parameters for SLobe are quite
different than for the SLobe1 region discussed above. The best-fit
photon index of the power-law model fit to the SLobe spectrum,
$\Gamma=2.73^{+0.24}_{-0.23}$, is larger than the one evaluated for
SLobe1. Not surprisingly, in the case of the thermal model, the
best-fit temperature for the SLobe spectrum is significantly lower
than the one obtained for SLobe1, namely $kT_{\rm b} = 0.8 \pm 0.2$\,keV.
These results may indicate that in the observed X-ray emission in the
SLobe feature is purely thermal in origin. Yet, the derived gas
temperature is slightly higher than that of the analogous structure to
the North (discussed below).

\subsubsection{Northern Hotspot}

The best-fit photon index in the power-law model for the X-ray feature
NHspot was determined as $\Gamma=3.34^{+0.16}_{-0.15}$, assuming
Galactic absorption only. Its spectrum is therefore much steeper than
that determined for SHspot. The spectrum appeared even steeper
($\Gamma=5.0^{+0.1}_{-0.1}$) when the absorption column was allowed to
vary during the fit, returning $N_{\rm H} = 3.32^{+1.12}_{-1.00}
\times 10^{21}$\,cm$^{-2}$ well in excess of the Galactic value, but
significantly less than the column measured for the nucleus. These
results may imply either a non-negligible intrinsic absorption of the
non-thermal X-ray continuum of NHspot, possibly related to the
presence of line-emitting gas in the vicinity of the hotspot, or a
thermal origin of the X-ray emission of the hotspot.  In the framework
of the thermal scenario we obtained a modest temperature of the
emitting plasma $kT_{\rm b} =0.54^{+0.06}_{-0.04}$\,keV and
$0.49^{+0.07}_{-0.06}$\,keV for both frozen and free absorption
column, respectively. (We note that in the latter case, the intrinsic
absorption $N_{\rm H} = 7.8^{+6.7}_{-5.8} \times 10^{20}$\,cm$^{-2}$
turned out anyway to be consistent with the Galactic value.) The soft
($0.5-2$\,keV) and hard ($2-10$\,keV) luminosities of NHspot
(corrected for absorption) are $4.5 \times 10^{40}$\,erg\,s$^{-1}$ and
$1.5 \times 10^{39}$\,erg\,s$^{-1}$, respectively. Even though
spectrum has only 242.7$\pm15.8$ net counts in $0.5-7$\,keV band we
were able to fit the abundances (for the absorption fixed at the
Galactic value) and obtained the best fit value of
$A=0.06^{+0.03}_{-0.02}$ with a slightly higher value of the
temperature, $kT_{\rm a}=0.79^{+0.06}_{-0.06}$\,keV. We note that in
this model these two parameters, $N_H$ and $A$, are correlated and the
true uncertainties on both parameters are larger.  In addition our
very low counts spectrum would not resolve any multi-temperature
structures resulting in the artificially low abundances.

NHspot is the brightest X-ray feature in the northern region being
embedded within the defined larger-scale NLobe region where the soft
X-ray emission is rather diffuse. We modeled the NLobe spectrum with a
thermal model and obtained $kT_{\rm b} = 0.45^{+0.05}_{-0.04}$\,keV
when we apply Galactic absorption only and a slightly lower
temperature $kT_{\rm b} = 0.33^{+0.09}_{-0.08}$\,keV with a relatively
strong absorption of $N_{\rm H} = 1.57^{+0.91}_{-0.08} \times
10^{21}$\,cm$^{-2}$.  The NLobe temperature is in good agreement with
the temperature estimated above for the NHspot which strongly supports
a thermal origin of the X-ray emission from the hotspot.  We note
  that the abundances are low 0.06$^{+0.03}_{-0.02}$ and consistent
  with the value found for NHspot above. The $0.5-2$\,keV unabsorbed
luminosity of the entire NLobe region assuming the thermal model is
equal to $7.5\times 10^{40}$\,erg\,s$^{-1}$, which is about twice the
luminosity of NHspot in the same band.

We also modeled the spectra of the region which corresponds to the
outermost northern enhancements visible on the edges of the radio
lobe, NLobe1 (see Figure~\ref{lobe-narrow-band}).  We obtained a
significantly higher temperature ($kT_{\rm b} =1.9^{+1.0}_{-0.5}$\,keV) for
NLobe1 than for the other parts of the northern X-ray emission. It is
also interesting to note that a power-law model fit to the X-ray
spectra of NLobe1 gave $\Gamma=2.16\pm0.28$, which agrees with the
photon indices of SHspot and also SJet. Thus this emission may
be of a different origin than the emission from NLobe and NHspot and
be more closely related to the radio outflow and interaction with the
ambient medium. We continue our discussion of the northern emission in
Section~\ref{sec:discussion} below.

\subsection{Diffuse Emission}

We identified several distinct X-ray emission regions across the
galaxy in addition to the main morphological features shown above.
They are indicated in Figure~\ref{spec-regions} as EArm, HBranch, and
HBranchS.  These regions were modeled with both an absorbed power-law
model and a thermal emission (see Tables~\ref{table-4} and
\ref{table-5}, respectively). The best-fit model parameters for the
EArm and HBranch regions are identical, with $\Gamma = 3.53$ for the
power-law and $kT_{\rm b} = 0.39$\,keV for a thermal model fit. 
  The metal abundances are similar for both regions with
  $A=0.13-0.18$ of the Solar value for the temperature of $kT_{\rm a} =
  0.7$\,keV.  The HBranchS spectrum has a slightly higher best-fit
  temperature of $kT_{\rm b} = 0.52^{+0.06}_{-0.05}$\,keV and lower but
  consistent metal abundance of 0.09$^{+0.08}_{-0.04}$ for
  $kT_{\rm a} =0.75$\,keV.  The total $0.5-2$\,keV luminosities (corrected for
absorption) of all three components are relatively high and equal to
$9.1 \times 10^{41}$\,erg\,s$^{-1}$. The total contribution of these
diffuse components to the total emission in the hard X-ray band is at
least 3 orders of magnitude lower. The thermal origin of these regions
is advocated in Section~\ref{sec:discussion} below.

\section{Discussion}
\label{sec:discussion}

We have presented new, deep \chandra\ X-ray observations of the radio source,
4C+29.30, hosted by a nearby ($z=0.0647$) post-merger elliptical galaxy.
The double-double radio morphology of this system consists
of Mpc-size relic lobes and a smaller-scale, younger radio source
\citep[][Section~\ref{sec:intro}]{jamrozy2007}.  Our X-ray observations focused on
the smaller-scale (inner tens of kpc-size) radio structures where we
could study the effects of interactions between the expanding radio
source and the ISM.

4C+29.30 could be viewed as an analogue to the famous radio galaxy
Centaurus~A \citep{israel1998}.  They display a similar morphology
with a pronounced dust lane and cold gas situated across the galaxy
perpendicular to the radio source axis, but the former is slightly more
powerful at radio frequencies than Centaurus~A.  4C+29.30 is also
characterized by a luminous extended emission-line region which is of
the same size ($\sim 50-60$\,kpc) as the inner radio source.
\citet{vanbreugel1986} studied a connection between the radio source
and line-emitting gas using VLA and ground-based optical data. They suggested
that the gas is most likely dragged from the inner parts of the host
galaxy, accelerated, and partly entrained by the radio-emitting plasma
outflowing from the nucleus. The process responsible for the
ionization of the line-emitting gas far away from the nucleus remained
vague however.

Our \chandra\ X-ray observations of 4C+29.30 add a new dimension to
studies of the physics of radio source interactions with the ISM in
general.  The new deep X-ray dataset enabled us to determine the
overall energetics of the system with much improved detail over the
earlier studies in the radio and optical bands. The \chandra\ image
revealed a remarkable X-ray morphology on the scale of the young radio
source and the extended emission-line region, with complex diffuse
filaments and compact blobs partly overlapping with the radio-emitting
plasma, and partly with the line-emitting gas.  The spectral analysis
carried out for several of the brightest X-ray features indicated a
mixture of thermal and non-thermal emission components, characterized
by a variety of temperatures and spectral slopes. We also detect
  possible variations in metal abundances of the hot ISM with the high
  abundances in the center and the very low values ($<0.1$ of the
  Solar) in the outer regions of the galaxy. However, the quantitative
  measurements of the abundance variations are not possible with the
  current data. The presented results support and strengthen the
conclusions regarding the feedback process operating in 4C+29.30,
which we discuss below.

\subsection{Nucleus of 4C+29.30}
\label{sec:agn}

Even though the nucleus of 4C+29.30 was detected in X-rays already in
the previous short \chandra\ observation \citep{gambill2003}, only now
can we study its properties in details using the new high-quality
spectrum at keV photon energies.  The new observations confirmed that
the AGN is buried within a large amount of gas and the AGN emission is
only visible at the high energy X-rays where the effect of the strong
absorption is less substantial.  The measured intrinsic absorption
column of $N_{\rm H} = 3.95^{+0.27}_{-0.33} \times 10
^{23}$\,cm$^{-2}$ is consistent with the values reported earlier by
\cite{gambill2003} and \cite{jamrozy2007}.  This X-ray absorption is
due to both the neutral and ionized gas.  The neutral Hydrogen column
density of the in-falling gas based on the redshifted HI absorption
lines reported by \citet{chandola2010} is equal to $N_{\rm HI} \simeq
5 \times 10^{21}\, (T_{\rm s} / 100\,{\rm K})$\,cm$^{-2}$, where
$T_{\rm s}$ is the spin temperature.  The standard assumption of
$T_{\rm s} \sim 100$\,K results in almost two orders of magnitude
smaller neutral column than the total one measured in X-rays, i.e.,
$N_{\rm HI} \ll N_{\rm H}$.  Only with much larger spin
temperatures (which might in fact be expected in the AGN environment)
could the neutral Hydrogen column density around the nucleus of 
4C+29.30 be comparable to the total equivalent
Hydrogen column density inferred from our X-ray observations
\citep[see in this context the discussion in][]{ostorero2010}. Note
that large amounts of in-falling matter inferred from the redshifted
HI absorption lines are usually observed in rejuvenated and newly born
radio sources \citep[e.g.,][and references
therein]{gupta2006,saikia2007}, including also the case of
Centaurus~A \citep{morganti2008}.

The nuclear X-ray luminosity of 4C+29.30 is equal to $L_{\rm 2-10\,keV}
\simeq 5 \times 10^{43}$\,erg\,s$^{-1}$, a value typical of nearby
Seyfert-type AGNs \citep{singh2011}. This, together with the best-fit
photon index for the hard X-ray nuclear continuum ($\Gamma \simeq
1.7$; representative of Seyferts and radio galaxies), suggests that
the observed radiation is related to the standard AGN engine, i.e.,
X-ray emission from a hot corona reprocessing thermal optical/UV
radiation from an accretion disk \citep[see
e.g.,][]{chiang2002,sobolewska2004a,sobolewska2004b}.  In the case of
4C+29.30, the exact level of the accretion-related UV luminosity is
unclear because of the significant absorption affecting the observed central
optical/UV radiation.  However, assuming a typical value of the
UV--to--X-ray luminosity ratio found in AGN samples corresponding to a
power-law slope of $\alpha_{\rm ox} \equiv -0.3838\times \log[L_{\rm
  2\,keV}/L_{\rm 2500\,A}] \simeq 1.5$
\citep{kelly2007,sobolewska2009}, one can evaluate the optical/UV
luminosity of the 4C+29.30 nucleus to be roughly $\sim
10^{45}$\,erg\,s$^{-1}$. On the other hand, if we assume a bolometric
correction factor of 70 for Seyfert 2 galaxies, as advocated by
\cite{singh2011}, we obtain a bolometric luminosity of 3.5$\times
10^{45}$\,erg\,s$^{-1}$.

We note that 4C+29.30 has recently been detected at photon energies
exceeding 10\,keV and up to $\sim$150\,keV by the \textit{Swift}/BAT
\citep{baumgartner2010}. The photon index of a power-law model
reported in the \textit{Swift}/BAT 58-month
catalog\footnote{{\texttt http://heasarc.nasa.gov/docs/swift/results/bs58mon/}}
of $\Gamma_{\rm BAT}= 1.65^{+0.45}_{-0.42}$ agrees with the photon
index we found in our \chandra\ spectra of the nucleus. We note that
the reported \textit{Swift}/BAT luminosity in $14-150$\,keV energy range is
equal to $1.7\times10^{44}$\,erg\,s$^{-1}$. This emission is most
likely related to the disk corona in the Seyfert-type core of
4C+29.30. This provides additional evidence for the intrinsic high
luminosity of the AGN buried in the 4C+29.30 center as we concluded
based on the new \chandra\ data.

Our estimated AGN luminosity is comparable to the accretion-related
luminosities observed in quasars \citep{hardcastle2009}, and therefore
exceeds by orders of magnitude the non-stellar nuclear optical-UV
continuum in 4C+29.30 estimated by \citet{vanbreugel1986}. Yet it is
in agreement with the required photoionizing flux for the observed
equivalent width (EW) of the [OIII] emission-line in the extended
emission-line region (a total required luminosity of the order of
$\sim 10^{43}$\,erg\,s$^{-1}$).  \citet{vanbreugel1986} excluded
collisional ionization of the line-emitting gas given the observed
line ratios and speculated on the source of the ionizing photons. Our
new results imply that the AGN can provide the required energy to
efficiently photo-ionize the extended emission-line region gas.
Further investigations of the nature of the AGN buried in the center
of 4C+29.30 are given in \citet{sobol2012}.

\subsection{Broad-band Emission of the Jet}
\label{jet}

\begin{figure*}
\begin{center}
 \includegraphics[width=\columnwidth]{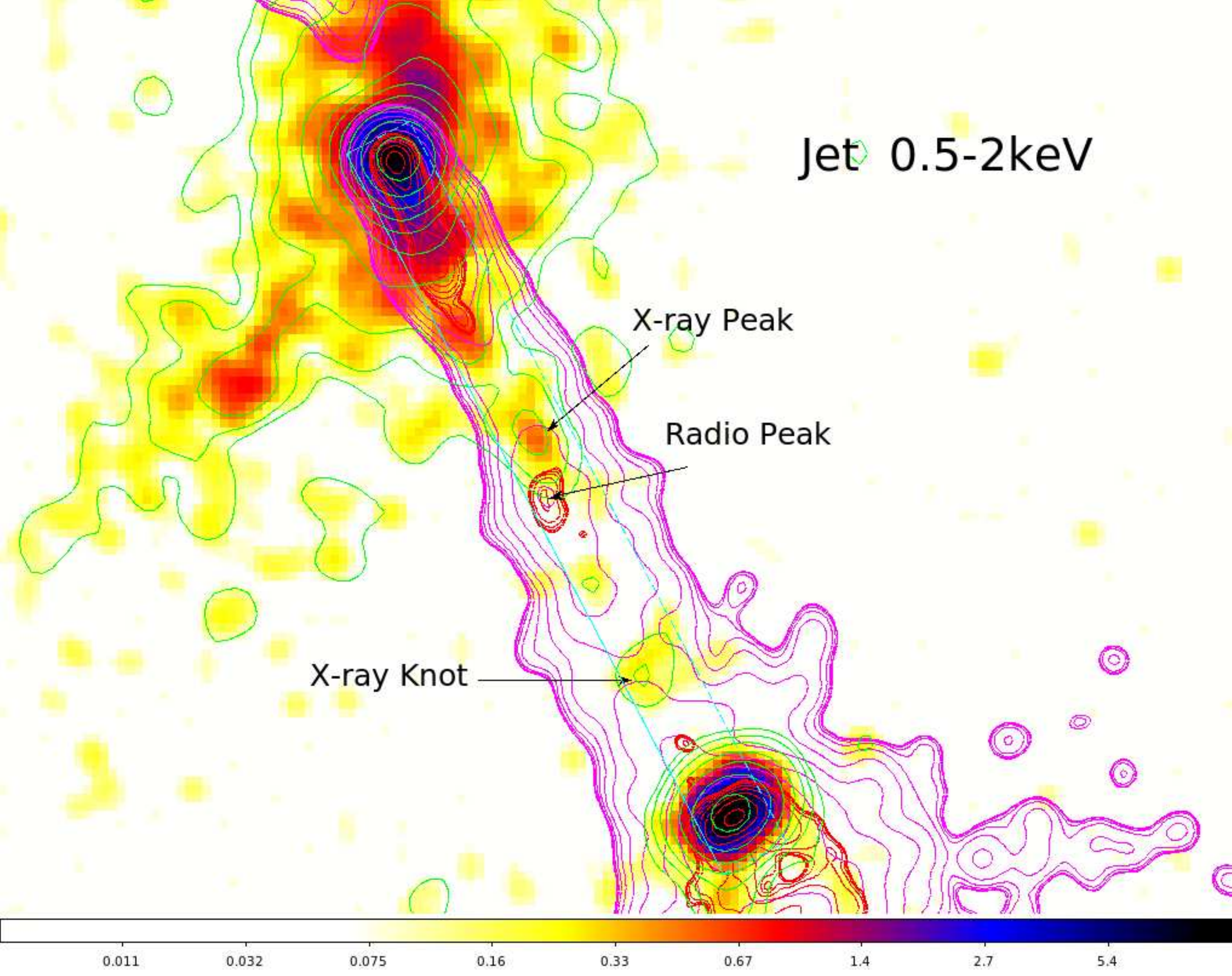}
 \caption{\small X-ray image ($0.5-2$~keV) of the 4C+29.30 southern region
   where the radio jet is observed.  Green contours mark X-ray emission, magenta
   contours show 1.4~GHz emission and red contours mark 5~GHz emission based on
   1\arcsec\ and 0.5\arcsec\ resolution maps, respectively. Cyan marks the box
   regions assumed for the extraction of the jet spectrum. We mark the
   positions of X-ray and radio peaks in the map corresponding to the
   features in the profiles. Both radio images were reprocessed by
   \citet{cheung2004} from archival VLA data published by
   \citet{vanbreugel1986}.}
 {\label{Xradio-contours}}
\end{center}
\end{figure*}

The one-sided radio jet of 4C+29.30 extends from the core to the south
beyond the host galaxy and portions of it were detected in our deep
\chandra\ image. Excluding the hotspot and core regions, a total radio
luminosity of the jet is roughly $\sim 3 \times
10^{40}$\,erg\,s$^{-1}$ \citep{vanbreugel1986}. The X-ray spectrum is
best modeled by an unabsorbed power-law continuum with a relatively
large photon index of $\Gamma=2.2\pm0.2$ resulting in a total X-ray
luminosity of $\sim 3 \times 10^{40}$\,erg\,s$^{-1}$, i.e., comparable
to the jet radio luminosity.  Therefore the observed X-ray emission is
most likely non-thermal in origin, although thermal scenarios cannot
be definitively excluded. That is because the jet X-ray emission is
relatively faint, and hence we were forced to extract the spectrum
over a relatively large extraction region.  This average spectrum may
be over-simplified since there are several `blobs' or `patches' of
diffuse X-ray radiation scattered across the entire galaxy, and there
is a non-negligible probability that a few such blobs would fall into
this extraction region. Indeed, the jet's X-ray morphology consists of
several isolated knots, and it is unclear whether there is any
continuous low-surface brightness emission associated with the
inter-knot regions (see Figure~\ref{Xradio-contours}). The only sign
of possible continuous X-ray emission aligned with the radio jet can
be found in the vicinity closest to the core. However, there is a
similar X-ray feature extending on a similar scale in the opposite
direction from the core (to the north) and both features coincide with
the known emission-line filaments which indicate the presence of a
warm medium \citep{vanbreugel1986}.

We note that our \chandra\ image may not be deep enough for a
detection of the continuous X-ray jet emission. For example, in the
case of the X-ray jet of Centaurus~A \citep{hardcastle2003}, there are
many surface brightness enhancements embedded in a \emph{faint}
diffuse continuous emission (less than $\sim 10\%$ of the knots'
peaks).  Many of these X-ray enhancements (though not all) have
corresponding radio knots, and some of them also show offsets between
the radio and X-ray peak brightnesses.  In 4C+29.30, the limits to the
inter-knot X-ray emission are of the order of $30-40\%$ of the knots'
peaks.  It is then possible that the jet X-ray diffuse emission is
present but just below our detection limits and we are only able to
detect X-ray knots.

We present a zoomed-in X-ray image of the jet overlaid with the radio
contours at two different frequencies in
Figure~\ref{Xradio-contours}.  The lower-resolution $1.4$\,GHz radio
map shows a continuous and slowly expanding outflow flaring at the
position of the southern hotspot. The higher-resolution 5\, GHz radio map 
reveals instead a very knotty structure, with the innermost
$4.3''$-scale (or 5\,kpc-long) linear feature terminating at the
position of the X-ray enhancement located at the boundary of the
$1.4$\,GHz jet.  An isolated 5\,GHz knot is present further away
($\sim 8\arcsec$ from the core), and \emph{downstream} of the
strongest X-ray peaked bright feature.  Some faint enhancements in the
X-ray emission are seen all along the radio jet further downstream up
to the terminal hotspot where both the radio and X-ray emission peaks
are well-aligned.

We cannot exclude some significant thermal contribution to the jet
X-ray emission spectrally given the limited quality of the current
data.  However, in the case of a non-thermal origin for the bulk of
the detected jet X-rays, two main processes can be invoked --
synchrotron emission from high-energy relativistic electrons and the
inverse Compton (IC) process involving low-energy particles only.
There is strong evidence for the synchrotron origin of the X-ray
emission in all FR\,I jets detected by \chandra\ \citep[see][and
references therein]{harris2006} in general, including the Centaurus~A
radio galaxy \citep{kataoka2006,worrall2008,goodger2010}.  The
synchrotron process has also been advocated for the X-ray jets in some
FR\,II sources, such as Pictor\,A and 3C\,353
\citep[e.g.,][]{hardcastle2005,kataoka2008}.  The X-ray photon index
and luminosity, and the radio--to--X-ray flux ratio inferred for the
4C+29.30 jet are similar to those established for the other FR\,I
objects, supporting the synchrotron interpretation also in this case.

In the synchrotron model, the Lorentz factors of the X-ray emitting
electrons in the jet need to be as high as $\gamma \simeq {1 \over 2}
\, \delta^{-1/2} \, (\nu/10^{18}\,{\rm Hz})^{1/2} \, (B/\mu{\rm
  G})^{-1/2} \sim 10^8$, for the equipartition magnetic field, $B
\simeq 30$\,$\mu$G, and the expected jet bulk Doppler factor, $\delta
\sim$ a few.  \citet{stawarz2002} argued that 100\,TeV-energy
electrons could be produced by turbulent acceleration within the whole
jet volume but especially at the jet boundaries where the interactions
between the radio jet and the ISM can generate substantial turbulence
\citep{deyoung1986}.  There is strong observational evidence for such
interactions to take place in 4C+29.30 and in the other low-power FR\,I 
systems in general. Electrons accelerated in this way could then be 
compressed by shocks at the positions of the knots.

Although, the synchrotron process is likely responsible for the jet
X-rays, the IC scenario, favored by many authors in the case of
powerful and X-ray-bright quasar jets
\citep{tavecchio2000,celotti2001}, remains a formal possibility.
There are several sources of seed photons for the IC scattering
located at tens of kpc distances from the nucleus: (1) the cosmic
microwave background (CMB), (2) starlight of the elliptical host
\citep{stawarz2003,hardcastle2011}; and (3) the beamed emission of the
misaligned blazar core \citep{celotti2001,migliori2011}. The relevance
of different photon fields depends on the velocity structure of the
jet, while the resulting X-ray IC spectrum reflects the energy
spectrum of the  low-energy tail of the electron distribution
\citep{harris2006}. In this
regard, one challenge to the IC hypothesis is that the observed
spectral index of the jet-related X-ray emission in 4C+29.30 is larger
than the spectral index of the jet radio continuum \citep[$\alpha_{\rm r} 
\simeq 0.8$; see][]{vanbreugel1986,jamrozy2007}. This is because
the higher-energy electrons involved in producing synchrotron radio
emission are not expected to be characterized by a flatter spectrum
than the low-energy electrons inverse-Compton upscattering
starlight/nuclear photons to the X-ray frequency range.

We conclude that the synchrotron scenario for the jet X-ray emission
is the most likely option, even though some contribution due to the IC
scattering of the starlight or of the blazar emission illuminating the
jet from behind, or even a thermal origin of some fraction of the
detected X-ray flux, cannot be excluded. A more in-depth understanding
of the broad-band emission of various jet features in the source would
however require more detailed modeling than considered here.

\subsection{Hotspots and the Structure of the Outflow}

\begin{figure*}
\begin{center}
  \includegraphics[width=5.5in]{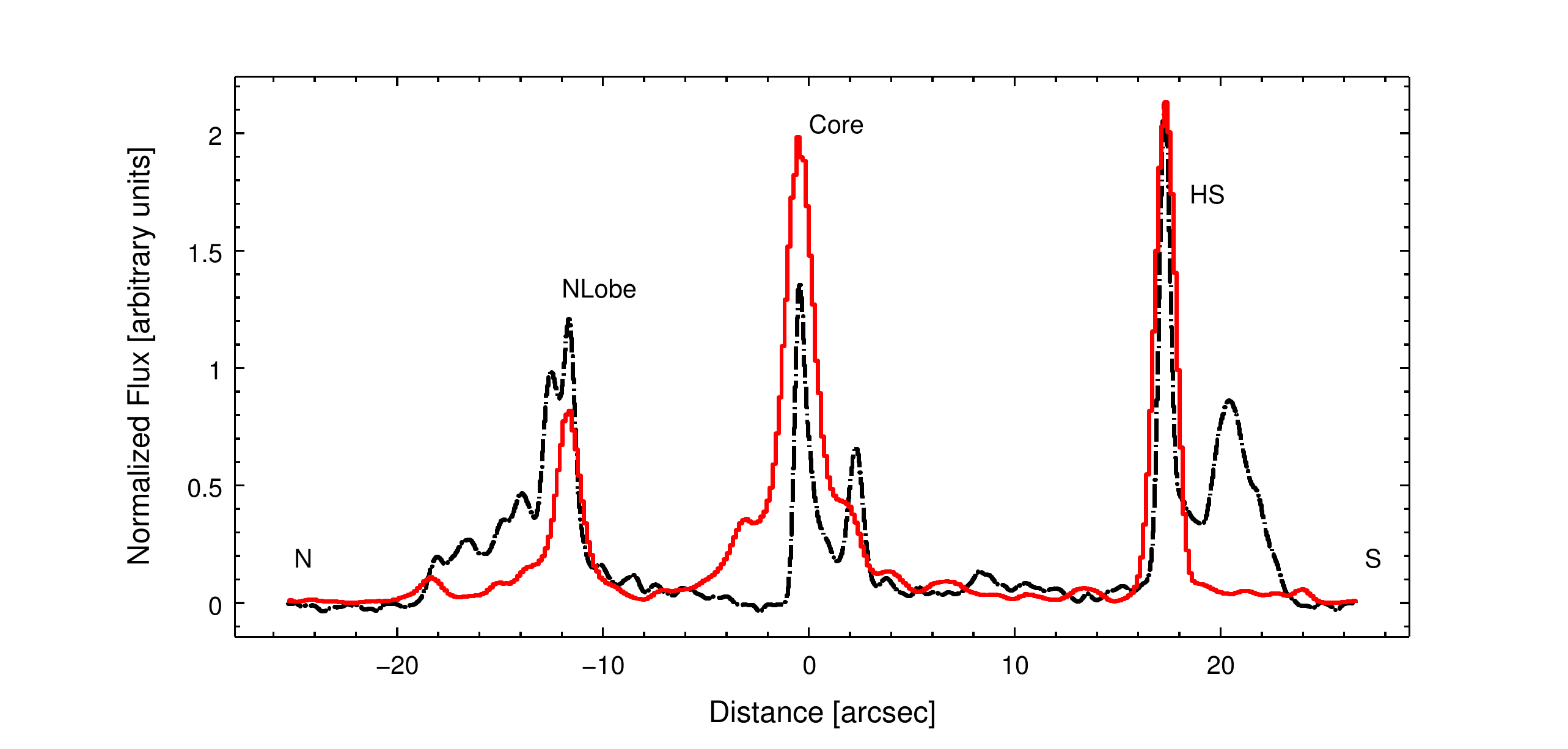}
  \caption{\small X-ray (0.5-2\,keV) and radio (5\,GHz) profiles
    extracted from a 4\arcsec\ wide box aligned with the jet at
    P.A.=24\deg\ (north is to the left and south to the right). The
    radio 5\,GHz profile is plotted with a dashed line and the X-ray
    profile is plotted as solid (red) line. The profiles are aligned
    at the core and normalized to the brightest peak to best compare
    the morphology in two bands.}
  {\label{profile}}
\end{center}
\end{figure*}

In Figure~\ref{profile} we show the radio and X-ray profiles along the
radio source axis (P.A. = 24\deg; here, the bulk of the AGN core
emission is absorbed, and hence the X-ray core peak in the figure is
dominated by the soft unabsorbed component). The emission at the two
frequencies appears correlated, with the brightest radio and X-ray
peaks aligned.  The characteristic double structures of the radio
features noted first by \cite{vanbreugel1986} for the brightest knots
in the radio jet and the hotspot regions can easily be seen in the
radio profile.  Note that the brighter peaks in these doubles are
always located upstream of the outflow and that the X-ray counterparts
have only been detected for these upstream features. In the observed
profiles, the radio flux asymmetry between the prominent southern
hotspot and the counter-hotspot to the north is also evident and this
asymmetry is followed in the X-rays.

Positional offsets between X-ray and radio peaks (with the X-ray peaks
located upstream relative to the radio ones) and double structures of
radio knots in general, have been seen in large-scale jets observed by
\chandra, e.g., the cases of Centaurus~A \citep{hardcastle2003},
PKS\,1127$-$145 \citep{siemiginowska2007}, and 3C\,353
\citep{kataoka2008}.  The origin of such offsets is still debated.
\citet{kataoka2008} suggested that such a morphology cannot be linked
to stationary jet features (like nozzles of reconfinement shocks, for
example), but instead could be due to moving portions of the plasma
generated by the central engine during epochs of enhanced activity.
\citeauthor{kataoka2008} further speculated that a complex
double-shock structure may form as a result of the interaction between
the two phases of the ejecta, with the reverse shock (propagating
within the faster portion of the jet) being associated with the peak
of the X-ray emission.  This scenario may also possibly account for
the properties of the 4C+29.30 jet and hotspots. If so, it is also of
particular relevance for understanding the intermittent/modulated jet
activity in this source \citep[see discussion in][]{jamrozy2007}.

 The non-thermal (synchrotron, in particular) character of the X-ray 
emission of the southern hotspot advocated by \citet{tavecchio2005} and 
supported by our spectral analysis (Section~\ref{sec:spectra}) fits the 
above model well. However, a possible thermal origin of the X-ray emission 
associated with the northern hotspot (on the counter-jet side) would then 
be puzzling.
There have only been a few detections of thermal X-ray emission from
the hotspots of powerful radio sources. In these cases, the emission
was apparently enhanced by interactions between a radio jet and dense
ISM clouds \citep{ly2005,young2005}.  Indeed, the northern hotspot of
4C+29.30 coincides with the region of a significant amount of observed
line-emitting gas, so may be an example of such a relatively rare
phenomenon.  But then the question remains why the X-ray peak
brightness coincides with the upstream radio peak rather than with the
outer (downstream) one. On the other hand, the soft X-ray spectrum of
the northern hotspot may be alternatively described by the high-energy
tail of the synchrotron continuum shaped by severe aging due to
radiative losses of the underlying electron energy distribution. The
spectral asymmetry between the southern and the northern hotspots
could then possibly be related to the brightness asymmetry, and hence
accounted for by relativistic effects (difference in relativistic
beaming of the plasma at the reverse shocks on the jet and the
counter-jet sides, light-travel effects, etc.) in the framework of the
modulated jet activity scenario.

Even more likely, the differences in the local ISM between the
regions north and the south of the nucleus could lead to distinctive
source morphologies in these two directions of the outflow. The radio
jet extending towards the south implies a possibly less disturbed
outflow than the one to the north characterized by enhanced optical
emission features due to interaction between the radio source and the
dense medium \citep{vanbreugel1986}.  Such environmental differences
were observed in other galaxies, although very high signal-to-noise
data from deep \chandra\ observations were required to accurately
determine the nature of the X-ray emission as was performed for
Centaurus~A \citep{kraft2009,croston2009}.  Interestingly, the
  very recent numerical simulations by \citet{gaibler2011} confirm
  that interactions of bi-polar relativistic jets with the
  asymmetrically distributed multi-phase ISM can indeed result in an
  asymmetric morphology of the entire system, with the jet that
  propagates into the denser environment being restrained (at least
  for some time), and affecting at the same time the line-emitting
  clouds themselves in various ways. Future X-ray and other
  multi-waveband observations of 4C+29.30 are needed to fully
  understand the structure of the outflows and the two hotspots in the
  source.

\subsection{Diffuse X-ray Features}

In addition to the core, jet, and hotspots, our \chandra\ image reveals 
diffuse X-ray emission covering large parts of the galaxy and extending 
beyond the radio-emitting plasma.  Some structures correlate with the 
extended emission-line region while others might be related to a recent 
merger event. For example, the two X-ray filaments perpendicular to the 
radio axis and extending from the core toward the SE and NW directions 
(HBranch and HBranchS regions in Figure~\ref{spec-regions}) seem to be 
aligned with colder gas mixed with the dust seen in the archival optical 
images of the host galaxy, resembling the famous dust lane in the 
Centaurus~A galaxy \citep{israel1998}. On the other hand, these two 
filaments may also be related to the jet activity. In particular, they 
may be a signature of gas compressed by the expanding lobes to form a 
quasi-linear structure perpendicular to the jet axis. Similar structures 
has been seen in Cygnus~A \citep{wilson2006}, although the horizontal 
structures \citep[named `cold belts' by][]{wilson2006} were not as straight as 
observed in 4C+29.30.

The diffuse X-ray structures seen to the north of the nucleus overlap
with the clouds of line-emitting gas associated predominantly with the
boundaries of the counter-lobe \citep{vanbreugel1986}. Our spectral
analysis indicate that the X-ray emission of these diffuse components
is most likely thermal, with gas temperatures ranging from $\simeq
0.4$\,keV up to $\simeq 0.6$\,keV (see Table~\ref{table-5}), with the
possible exception of the southern regions located around the radio
lobe downstream of the hotspot, for which either a non-thermal
contribution or higher gas temperatures ($\simeq 0.8$\,keV) remain
valid options.

Given the normalization in the best-fit thermal model and the fixed
emitting volumes (assuming 1\,kpc depths for their third dimensions),
we estimated the densities of the hot X-ray emitting gas in different
regions (see Table~\ref{table-6}). These densities vary roughly
between $n_{\rm e} \sim 0.02$\,cm$^{-3}$ in the southern regions, and
$\sim 0.2$\,cm$^{-3}$ around the center and in the northern regions.
It is interesting to note that the denser material is located in the
regions with prominent optical line emission.  The estimated values
are consistent with the densities of hot gas found in other elliptical
galaxies located in poor groups \citep{mathews2003}.  4C+29.30 might
be located in such a group. We checked the NED for known galaxies
located within 20\,arcmin ($\sim$1\,Mpc) of 4C+29.30 and with the
velocities within $\pm 1000$\,km\,s$^{-1}$, so close enough to be
possibly associated with a group. Only 4 galaxies with known redshifts
met these selection criteria. One galaxy is closer than $\pm
500$\,km\,s$^{-1}$ and only $\sim 4.5$\,arcmin away, so possibly
indicative of a group.  Future observations are needed to confirm this
conclusion.

The average pressures of the hot X-ray emitting gas are, 
$p_{\rm th} \simeq 2 \, n_{\rm e} \, kT \lesssim
10^{-10}$\,dyne\,cm$^{-2}$ (see Table~\ref{table-6} for the pressure values 
of the individual features). This pressure is then comparable to/lower
than the minimum non-thermal pressures of the jet, hotspots and the
lobes (all estimated assuming energy equipartition between the
radio-emitting electrons and the magnetic field), as well as the
pressure of the line-emitting clouds calculated by
\cite{vanbreugel1986}, i.e., $\sim
10^{-10}-10^{-9}$\,dyne\,cm$^{-2}$. We note that the pressure
estimates for the line-emitting clouds are rather uncertain, and that
our calculations assumed one size for the depth of all the analyzed
regions, while \citeauthor{vanbreugel1986} used different scales for
different components. Also, the minimum pressure calculations do not
take into account the possible contribution of non-radiating particles
(cosmic-ray protons in particular) to the jet/lobes' pressure
\citep[see e.g.][]{dunn2004}.

We conclude that it is most likely that the expanding radio source in
4C+29.30 is slightly overpressured with respect to the ambient medium
(i.e., the hot ISM component), but it is likely in pressure
equilibrium with the denser clouds of the colder material emitting the
optical lines.  We also note that \citet{vanbreugel1986} found a
velocity structure of the gas along the jet with increasing velocity
towards the outer regions of the galaxy. The outermost gas is
outflowing with velocities of $\gtrsim 500$\,km\,s$^{-1}$ exceeding
the escape velocity for the host galaxy. The largest velocities (and
the broadest emission lines) were found at the boundaries of the radio
lobes and hotspots.  Thus it is natural to conclude that the entire
outflow is driven by the jets.

\subsection{Shocks}

The analysis of our X-ray data and the above pressure estimates also
indicate that not only the colder line-emitting gas but also the hot
X-ray emitting gas can be pushed out by the expanding radio-emitting
plasma.  The process may be accompanied by the formation of weak
shocks around the lobes, as witnessed in several other analogous
systems \citep[such as Centaurus~A, M~87, or NGC~1275;
e.g.,][respectively]{croston2009,million2010,fabian2011}. The shocks
should then compress and heat the surrounding X-ray emitting gas and
this may be the origin of the slightly hotter diffuse emission
observed around the southern lobe of 4C+29.30.

We can estimate the Mach number of a shock which could heat this
southern region. Assuming the standard shock jump condition, one gets
the temperature jump:
\begin{equation}
{T_+ \over T_-} = { \left[ 2 \, \hat{\gamma} \, \mathcal{M}^2 - 
\left(\hat{\gamma} - 1 \right) \right] \, \left[ \left( \hat{\gamma} - 1 
\right) \, \mathcal{M}^2 + 2 \right] \over \left( \hat{\gamma} + 1 
\right)^2 \, \mathcal{M}^2} \quad ,
\end{equation}
where $T_+$ and $T_-$ are the downstream and upstream gas
temperatures, respectively, $\mathcal{M} \equiv v / c_{\rm s}$ is the
Mach number (as measured in upstream region), and the adiabatic index
of the gas (assumed to be the same for the upstream and downstream
medium) is $\hat{\gamma}$ = 5/3, as appropriate for a non-relativistic
fluid.  For the observed temperatures of 0.5\,keV for the ambient
medium and 0.8\,keV for the hotter component, the resulting Mach
number reads as $\mathcal{M}=1.6$.  This is a very reasonable Mach
number and such weak shocks have been discovered in radio galaxies
located in clusters of galaxies, such as Hydra~A, Cygnus~A, and
Perseus~A \citep[][respectively]{nulsen2005,wilson2006, fabian2006}.
However, a much stronger shock was associated with the southern lobe
of Centaurus~A \citep[$\mathcal{M} \simeq 8$;][]{croston2009}.
  Also, slightly higher Mach numbers ($\mathcal{M} \simeq 4$) have
  been claimed for the shocks driven by the radio structures expanding
  in the group environments of a low-power FR\,I radio galaxy
  NGC\,3801 and a Seyfert galaxy Markarian~6, albeit at smaller scales
  \citep[$\lesssim 10$\,kpc;
  see][respectively]{croston2007,mingo2011}.

We can check the velocity of the shock since we determined the
temperature and density of the different regions in our X-ray data.
For a $kT = 0.5$\,keV, the sound velocity in the upstream medium is
$c_{\rm s} = \sqrt{\hat{\gamma} \, k T / \mu m_{\rm p}} \simeq
360$\,km\,s$^{-1}$, assuming a non-relativistic gas
($\hat{\gamma}=5/3$) and the mean molecular weight of the gas
$\mu=0.6$.  A $\mathcal{M} \simeq 1.6$ shock propagating within the
gas characterized by this sound velocity implies the radio lobes
driving the shocks have an expansion velocity, $v_{\rm exp} =
\mathcal{M} \times c_{\rm s} \simeq 580$\,km\,s$^{-1}$.  This is
in excellent agreement with the highest velocity measured for the
line-emitting gas $\gtrsim 500$\,km\,s$^{-1}$ and strongly supports
the idea that feedback is operating in this galaxy.

The implied shock velocity of $580$\,km\,s$^{-1}$ can be used to
estimate a size for the region influenced by the shock during the
source lifetime. Assuming the source dynamical age of $\sim33$\,Myr
\citep{jamrozy2007}, the calculated source size is consistent with the
extension of the SLobe ($\sim20$\,kpc). We conclude that during the
lifetime of the source, the whole SLobe region could indeed be
shock-heated.

\begin{figure*}
\begin{center}
  \includegraphics[width=5.5in]{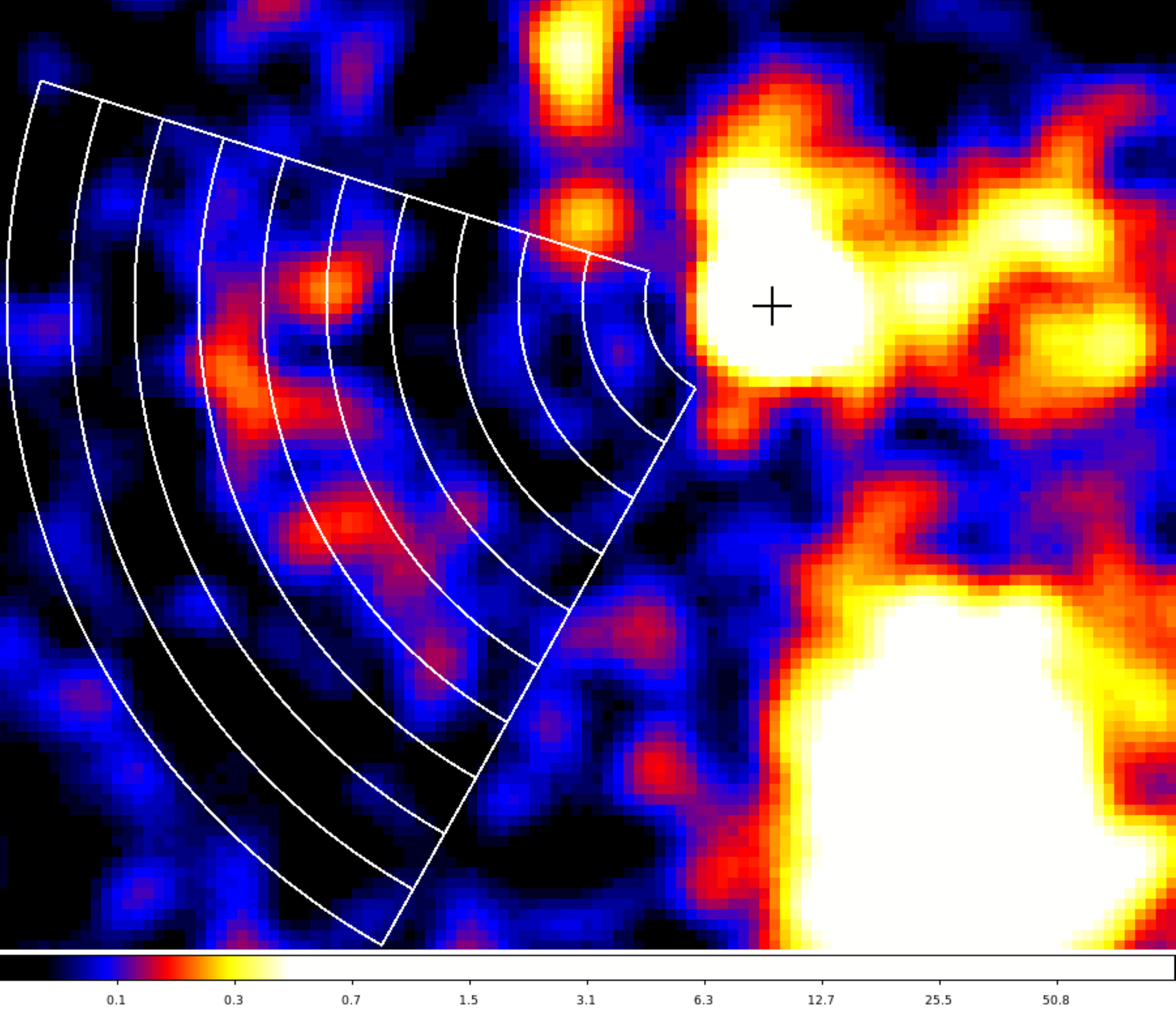}
  \caption{X-ray emission in the vicinity of NHSpot and EArm overlaid
    with regions (white) used to construct the surface brightness
    profile in Figure~\ref{earm}. Black cross marks the peak position
    of NHspot and the initial reference point for the regions.}
  {\label{fig:pie}}
\end{center}
\end{figure*}

\begin{figure*}
\begin{center}
  \includegraphics[width=5.5in]{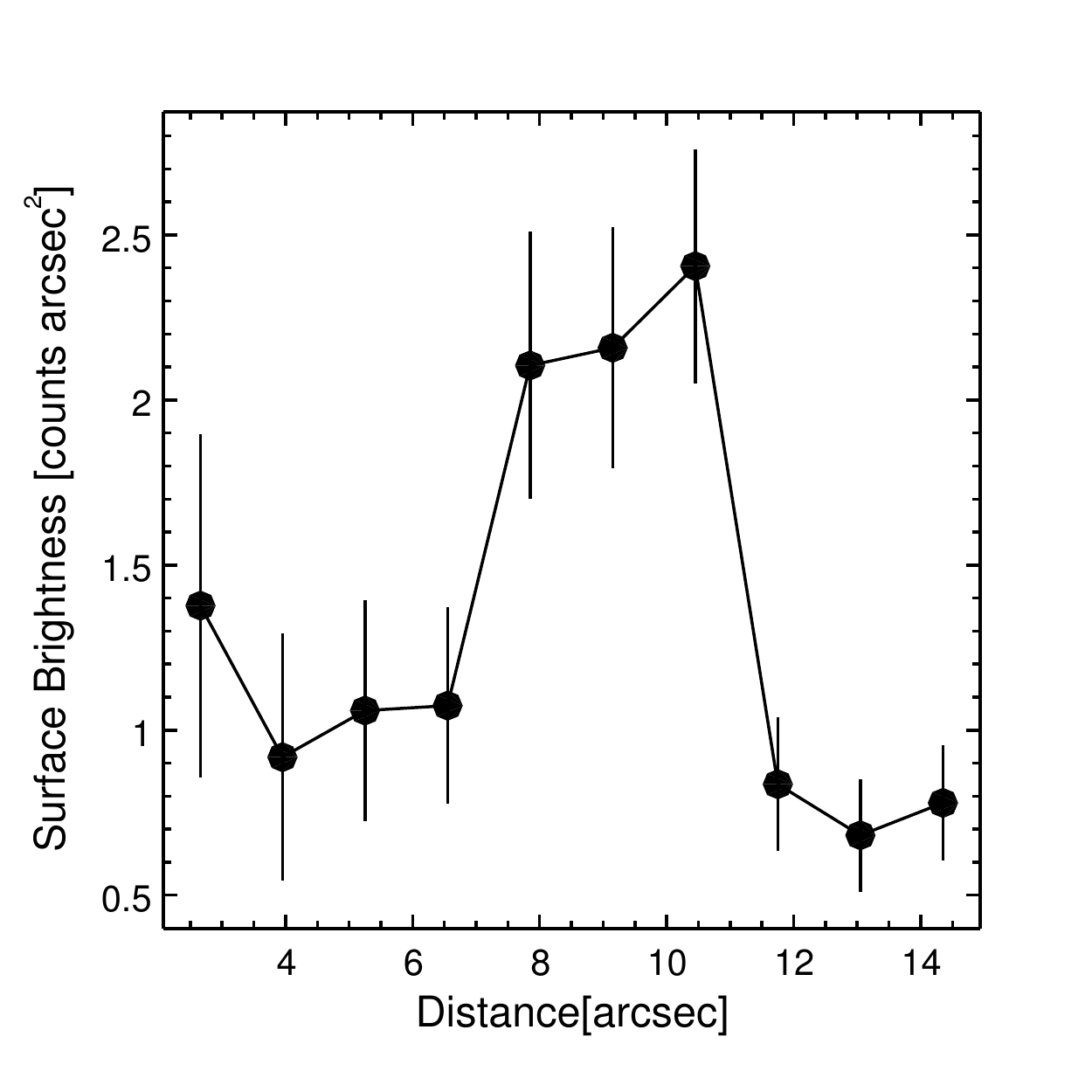}
  \caption{X-ray surface brightness profile (in units of
    counts\,arcsec$^{-2}$) across EArm constructed using the regions
    marked in Figure~\ref{fig:pie}. The distance is measured from the
    NHspot peak brightness marked in the region figure with black
    cross. }
  {\label{earm}}
\end{center}
\end{figure*}

We also comment on the northern X-ray emission region we identified as
EArm. This structure is located outside and along the eastern boundary
of the northern radio lobe along and is slightly offset from the
strongest radio emission.  Interestingly, \cite{vanbreugel1986}
reported a large velocity jump (by $\sim 300$\,km\,s$^{-1}$) in the
optical-line emitting gas in this region. In an X-ray surface
brightness profile across EArm (Figures~\ref{fig:pie} and \ref{earm}),
we found that the surface brightness increases by a factor of $\sim
2-2.5$ at the location of EArm over the projected width of
$\sim 3$\,kpc.  Although it is tempting to associate the surface
brightness increase as due to shock heating and compressing the
background plasma, we were unable to constrain the temperature profile
in this region in our current data.  However, we note that EArm could
be similar to the X-ray emission observed at the boundary of the
southern lobe of Centaurus~A \citep{croston2009} where the evidence
for the shock exists in a deep \chandra\ image.  The shock in
Centaurus~A is resolved on parsec scales (1\arcsec $\sim 18 $\,pc)
and the change in the surface brightness is measured across a $\sim
1$\,kpc region.  The EArm `width' is much larger, and the enhancement
covers about 3\,kpc, but its morphology seems to relate this structure
to the radio lobe.  Future higher signal-to-noise data are necessary
to properly assess the presence of a shock in the EArm feature.

We conclude that signatures of weak shocks due to an expanding radio
source are present in the X-ray data. The radio-emitting outflow in
4C+29.30 interacts therefore clearly with the ambient medium, not only
accelerating denser and colder optical-line emitting clouds present
within the ISM as found before, but also heating and displacing the
warm phase of its gaseous environment, thus providing strong evidence
for a complex jet-related feedback in a post-merger galaxy. We note
that interactions of a relativistic magnetized jet with a
\emph{multi-phase} ISM are rarely subjected to detailed numerical
simulations \citep[but see][]{krause2007,sutherland2007,wagner2011},
even though the problem is particularly relevant in the context of
high-redshift ($z > 1$) radio galaxies.

\subsection{Source Energetics}

The intrinsic luminosity of the nucleus of 4C+29.30 of
$\sim 10^{45}$\,erg\,s$^{-1}$ implies an accretion power of order
$\sim 10^{46}$\,erg\,s$^{-1}$, assuming a standard value of $10\%$ for
the radiative efficiency. It also implies a mass of the central 
supermassive black
hole that exceeds $\sim 10^8 \, M_{\odot}$. The observed non-thermal X-ray
luminosities of the jet and hotspots  put a strict lower limit on the
source radio power of $> 10^{41}$\,erg\,s$^{-1}$.

The minimum jet power can also be calculated from the radio
measurements. The total energy of the southern radio lobe (tail)
calculated under the equipartition condition is equal to $E_{\ell}
\sim 10^{57}$\,erg.\footnote{We note that the value
  reported by \cite{vanbreugel1986} in their Table~4B, is 2 orders of
  magnitude higher, which we believe is an error. Here, we assumed
  40\arcsec\, for the third dimension in calculating the volume (i.e.
  the same as in \cite{vanbreugel1986}) and also estimated the
  equipartition magnetic field of 3.4\,$\mu$G for the observed flux
  density of 0.26\, Jy at 1.4~GHz.}  The north and south radio structures 
(excluding the southern tail) have similar minimum energy estimates
of $\sim 10^{57}$~erg. If we now take the age of the radio source,
$t_{\rm j} \sim 30$\,Myr \citep{jamrozy2007}, we can obtain a lower
limit to the jet kinetic energy required to power the radio lobes
of $L_{\rm j} \sim E_{\ell} / t_{\rm j} \sim
10^{42}$\,erg\,s$^{-1}$.  Interestingly, this is also the minimum
power needed for heating the surrounding thermal gas required by our
\chandra\ observations.  The thermal energy in the diffuse X-ray
emitting component is $E_{\rm th} \sim 10^{56}-10^{57}$\,erg, which is
also in agreement with the minimum energy estimates for the radio
structures given above and would require a continuous energy supply of
$\sim 10^{41}-10^{42}$\,erg\,s$^{-1}$ over the radio source lifetime.

A total ionized Hydrogen mass in the line-emitting clouds of 4C+29.30
estimated by \citet{vanbreugel1986} is about $\sim 10^6 \, M_{\odot}$
and their total kinetic energy is of order $E_{\rm kin} \sim
10^{54}$\,erg.  This is much smaller than the total energy deposited
by the jets within the lobes. Thus we can conclude that only a small
fraction of the available jet power is needed to drive the gaseous
optical outflow, while a much higher fraction ($>10\%$) of the total
jet power ($\sim 10^{56}$\,erg) is used to heat the surrounding medium
to the X-ray temperatures ($>0.1$\,keV).

We can also compare the energy stored in the outer lobes formed during
the previous outburst of the radio source to the current jet
power.  Approximating the outer lobes as a cylinder with a
height $h=640$\,kpc and a radius $r=h/4$, the source volume is $V=\pi r^2 h$.
Assuming the minimum energy magnetic field
$B=0.7\,\mu$G ($U_{\rm B}=B^2/8\pi$) and lifetime $t_{\rm out} \geq 200$\,Myr given
by \cite{jamrozy2007}, 
the required jet power is equal to $L_{\rm j} \sim 2 \,
U_{\rm B}\, V / t_{\rm out} \sim 10^{43}$\,erg\,s$^{-1}$.  This is a
lower limit to the jet power in the previous outburst as the duration
of the active phase might be shorter than $t_{\rm out}$. 
However, this jet power indicates that the
previous outburst of the jet activity might have been more powerful
than the current one in agreement with the powers derived for other
double-double radio sources.

\section{Summary and Conclusions}\label{sec:summary}

We presented the results of our new deep ($\simeq 300$\,ksec) 
\chandra\ X-ray observations
of the nearby radio galaxy 4C+29.30.  These X-ray data brought a new
dimension to studies of radio source impact on the ISM and enabled
a detailed analysis of the processes associated with the feedback.

The main results of our studies can be summarized as follows:

\begin{itemize}

\item The new \chandra\ image of 4C+29.30 revealed a complex
  X-ray morphology with different features on the same scale ($\sim 50$\,kpc)
  as the radio (inner) source and the extended optical emission-line region.
  This indicates complex interactions between radio-, optical line-, and
  X-ray-emitting plasmas.
\item The nucleus of the galaxy is surprisingly powerful in X-rays
  ($\sim 10^{44}$\,erg\,s$^{-1}$) and heavily obscured by a significant
  amount of matter in-falling onto the center (also indicated by HI
  absorption lines). This infall may be related to the feeding of the
  nucleus and triggering the current jet activity. The nuclear
  emission is sufficiently luminous to photo-ionize the whole extended emission
  line region. In this respect, contrary to estimates based on the optical
  continuum only \citep{vanbreugel1986}, no other source of ionizing
  photons or other processes of ionization is needed.
\item The X-ray emission of the jet and the hotspots seems to be
  particularly complex, with different and distinct emission
  components, both thermal and non-thermal (synchrotron and/or inverse-Compton).
  Although we favor the synchrotron scenario for the jet and the
  hotspots, more definitive statements would require detailed modeling
  and higher quality multi-wavelength data in the future.
\item A significant fraction of the jet energy (jet power, $L_{\rm j}
  \sim 10^{42}$\,erg\,s$^{-1}$) goes into heating the surrounding gas
  and the X-ray data support the heating of the ambient medium via
  weak shocks with the Mach number, ${\cal{M}} = 1.6$.  Only a small amount
  of the jet power is needed to accelerate clouds of colder material
  which are dragged along the outflow.
\end{itemize}

Our study provides evidence for the AGN impact on the host galaxy ISM
and feedback processes. We show that a relatively large amount of the
energy generated by an accreting SMBH is dissipated by the outflow and
used to heat the ISM.  A smaller fraction of the energy drives the
colder gas out from the central regions and also heats up the medium.
Both processes can lead to a limited fuel supply and starving the
SMBH, and finally stopping its growth.

\section*{Acknowledgments}

The authors thank the anonymous referee for insightful comments and
Ma{\l}gorzata Sobolewska, Ralph Kraft, and Dharam Vir Lal for fruitful
discussion.  \L .S. is grateful for the support from Polish MNiSW
through the grant N-N203-380336. M.J. was supported by MNiSW funds for
scientific research in years 2009-2012 under the contract No.
3812/B/H03/2009/36. Work at NRL is sponsored by NASA DPR S-15633-Y.
This research has made use of data obtained by the Chandra X-ray
Observatory, and \chandra\ X-ray Center (CXC) in the application
packages CIAO, ChIPS, and Sherpa.  This research is funded in part by
NASA contract NAS8-39073. Partial support for this work was provided
by the \chandra\ grant GO0-11133X and XMM-Newton grant NNX08AX35G.

{}


\begin{table*}
{\scriptsize
\noindent
{\caption[]{\label{table-1}
Chandra Observations}}
\begin{center}
\begin{tabular}{cccccrc}
\hline\hline
\\
OBSID & Exposure (ksec) & Date \\
\\
\hline
\\
11688 & 123.4  & 2010-2-19 \\
11689 &  75.5  & 2010-2-25 \\
12106 &  32.0  & 2010-2-18 \\
12119 &  55.5  & 2010-2-23 \\
\\
\hline\hline
\end{tabular}
\end{center}
}
\end{table*}

\input{table2.tex}
\input{table3.tex}

\input{table4.tex}
\input{table5.tex}
\input{table6.tex}


\begin{figure*}
\begin{center}
\includegraphics[width=6.5in]{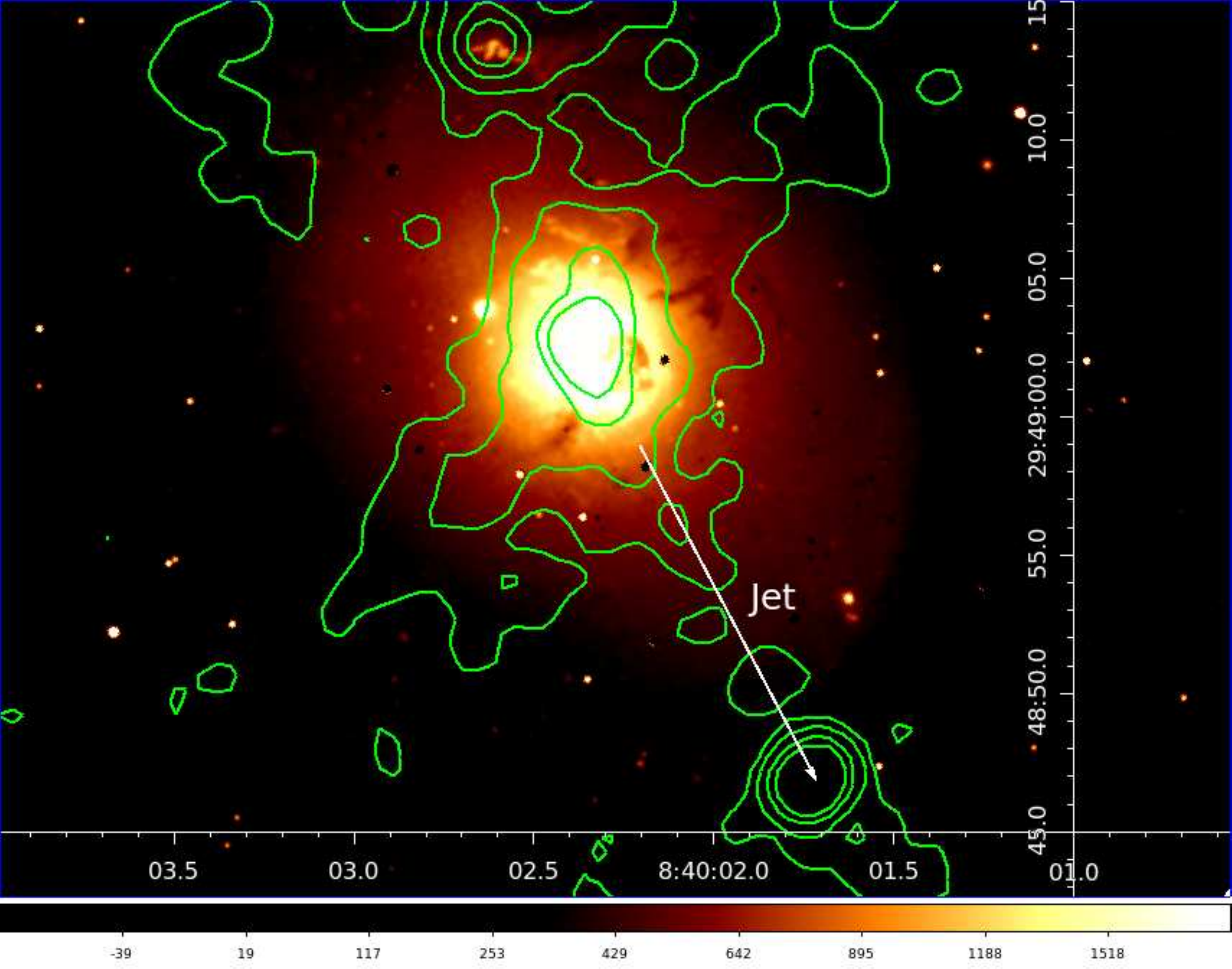}
\caption{Optical image of the host galaxy obtained from the HST
  archives (HST-STIS CCD, the wavelength peaks at 5840\AA). The colors
  represent intensity. The Chandra X-ray contours are overlayed on the
  image.  The SW direction of the radio jet is marked by a white arrowa.  Dark
  filaments of dust start in the censtral regions and extend
  perpendicular to the jet axis.}  {\label{optical}}
\end{center}
\end{figure*}


\end{document}

%% file: table2.tex
\begin{table*}
{\scriptsize
\noindent
\caption[]{\label{table-2} Spectral Regions.}
\begin{center}
\begin{tabular}{crrccccccccccc}
\hline\hline
\\
Name & Total Cts$^a$  & Net Cts$^a$ & Region Definition$^b$ \\
& 0.5-7\,keV & 0.5-7\,keV & \\
\\
\hline
\\
NLobe   & 319 & 275$\pm19$  & ellipse(08:40:02.34,+29:49:14.63,0.1135',0.07073',0)\\
        & & & -circle(08:40:02.61,+29:49:13.48,0.0267') \\
\\
NLobe1  &  75 & 62$\pm 9$  & rotbox(08:40:02.925,+29:49:17.42,0.05308',0.12527',0)\\
\\
NHspot  & 247 & 242$\pm16$ &  circle(08:40:02.613,+29:49:13.48,0.02673')\\
\\
SHspot	& 558 & 553$\pm 24$ & circle(08:40:01.729,+29:48:47.04,0.02907')\\	
\\
SJet   & 126 & 110 $\pm 12$ &rotbox(08:40:02.019,+29:48:54.20,0.041933',0.2033',31.7134)\\
\\
SLobe    & 301 & 169$\pm 21$ &  ellipse(08:40:02.110,+29:48:42.35,0.22604',0.102',12.9697)\\
	 & & &  -circle(08:40:01.729,+29:48:47.04,0.0292')\\

\\
SLobe1	  &145  & 97$\pm14$ &  circle(08:40:01.607,+29:48:43.77,0.0943')\\
	  & & & -circle(08:40:01.729,+29:48:47.04,0.0292')\\
\\

Center	  & 1097 & 1059$\pm 34$ & ellipse(08:40:02.341,+29:49:02.69,0.06631',0.10317',0)\\
	  & & & -circle(08:40:02.345,+29:49:02.61,0.021')\\

\\
Total	  & 3500 & 2574$\pm 66$ & rotbox(08:40:02.399,+29:49:01.14,0.635105',0.7768',0)\\
	  & & & -circle(08:40:02.345,+29:49:02.61,0.021')\\
	  & & & -circle(08:40:01.743,+29:48:46.95,0.0295')\\
	  & & & -circle(08:40:02.606,+29:49:13.66,0.0295')\\

\\
EArm	   & 124 & 92$\pm12$ & pie(08:40:03.00,+29:49:12.27,0.047682',0.1256',123.322,266.705)\\
\\
HBranch	   & 340 & 238$\pm21$&  rotbox(08:40:01.374,+29:49:14.02,0.122393',0.4380',311.166)\\
\\
HBranchS & 356 & 295$\pm 20$ &  rotbox(08:40:02.776,+29:48:55.52,0.19036',0.16845',38.496) \\
\\
Core	  & 5328 & 5303$\pm73$ & circle(08:40:02.345,+29:49:02.61,0.02083') \\
\\
Core\_ext   & 833 & 676$\pm32$  & annulus(08:40:02.345,+29:49:02.61,0.0246',0.0754') \\

\\
\hline

\end{tabular}

\smallskip

Note: $^a$ Total and background subtracted counts in each region in energies between 0.5\,keV and 7\,keV.

$^b$ CIAO regions  are given in J2000 and the arcmin units for the sizes.
\end{center}

}
\end{table*}

%% file: table3.tex
\begin{table*}
{\scriptsize
\noindent
\caption[]{\label{table-3} Best-Fit Model Parameters for the Nucleus.}
\begin{center}
\begin{tabular}{ccccccccccccccc}
\hline\hline
\\
Component &  Parameters$^a$ & Norm$^b$ & \\
&& [photons~cm$^{-2}$\,s$^{-1}$\,keV$^{-1}$] &  \\
\\
\hline
\\
&Nucleus model	--    gal(bremss+abs*pow) & \\
\\
\hline
\\
Absorption  &	 $N_H = 3.95^{+0.27}_{-0.33}\times10^{23}$\,cm$^{-2}$ & \\

Power Law &	 $\Gamma=1.70^{+0.38}_{-0.36}$ &  1.26$^{+1.48}_{-0.64}\times10^{-3}$\\

Bremstrahlung &   kT = 6.8$^{+9.9}_{-2.5}$~keV & 4.2$^{+0.5}_{-0.2}\times10^{-6}$ & \\

\\
\hline
\\
&Nucleus model  --	gal (apec+abs*pow) &\\
\\
\hline
\\

Absorption  &	 $N_H = 4.37^{+0.17}_{-0.30}\times10^{23}$\,cm$^{-2}$ & \\

Power Law & $\Gamma=1.87^{+0.36}_{-0.34}$ &  1.79$^{+1.87}_{-0.88}\times10^{-3}$\\

APEC$^c$   & kT = 9.6$^{+10.0}_{-4.9}$ & 1.5$^{+0.1}_{-0.3}\times10^{-5}$ & \\
& $A < 1.34$ & \\
\\
\hline 
\\
&Nucleus model  --     gal (pow\_soft+abs*pow) & \\
\\
\hline
\\
Absorption & 	 $N_H = 3.93^{+0.17}_{-0.30}\times10^{23}$\,cm$^{-2}$ & \\

Power Law &  $\Gamma=1.68^{+0.36}_{-0.34}$ &  1.21$^{+0.12}_{-0.60}\times10^{-3}$\\

Soft Power Law &  $\Gamma=1.60^{+0.12}_{-0.12}$ &  3.18$^{+1.87}_{-0.88}\times10^{-6}$\\

\\         
\hline
\\
& Background model -- scale*gal(bremss+pow) &\\
\\
\hline
\\
Power Law   &  $\Gamma=-2.3^{+0.2}_{-0.2}$ &  3.8$^{+1.6}_{-1.2}\times10^{-8}$ \\

Bremsstrahlung &  kT = 0.59$^{+0.04}_{-0.04}$~keV & 2.57$^{+0.27}_{-0.24}\times10^{-5}$ & \\

Scales$^c$  & 1.07$\pm0.07$ 1.15$\pm0.08$ 1.09$\pm0.09$ \\
\\
\hline
\end{tabular}
\end{center}

NOTE -- Components in model expression:
gal - galactic absorption $N_H$=3.98$\times10^{20}$\,cm$^{-2}$ - fixed in all models;
abs - intrinsic absorption of the nucleus; 
power law - absorbed intrinsic power law component, absorbed;
bremss - unabsorbed bremsstrahlung thermal emission;
apec - unabsorbed APEC plasma model.
$^a$ Uncertainties are 1$\sigma$ for one interesting parameter. 
$^b$ A model normalization is given at 1~keV.
$^c$ $A$ - an abundance parameter in APEC model expressed in the Solar values.
$^d$ Scale factors in respect to the data set 1.

}
\end{table*}

%% file: table4.tex


\begin{table*}
{\scriptsize
\noindent
\caption[]{\label{table-4} Parameters of Power Law Model Fits to X-ray Spectra of Various Regions.}
\begin{center}
\begin{tabular}{ccccccccccccccc}
\hline\hline
\\
Region &N$_H$$^a$ &	$\Gamma$  &Flux(0.5-2)$^b$ & Flux(2-10)$^b$ & Stat(Cash)$^c$ \\
\\
& [10$^{22}$\,cm$^{-2}$] &  &  [10$^{-15}$\,erg\,cm$^{-2}$\,s$^{-1}$] & [10$^{-15}$\,erg\,cm$^{-2}$\,s$^{-1}$] & \\
\\
\hline
\\
NLobe &  -- & 3.5$_{-0.2}^{+0.2}$ & 45.9$\pm 0.3$ & 0.64$\pm 0.03$ & 3527.26 \\
      & 0.41$_{-0.14}^{+0.15}$ & 5.9$_{-0.1}^{+0.1}$ & 31.6$\pm 15.0$ & 0.18$\pm0.15$ & 3518.0 \\
\\
NLobe1 & -- & 2.16$_{-0.28}^{+0.28}$ & 0.82$\pm 0.06$ & 0.75$\pm 0.41$ & 3082.28  \\
       & $<0.15$ & $<2.5$  & 1.0$\pm2.7$ & 0.9$\pm0.4$ & 3082.12 \\
\\
NHspot & -- & 3.34$_{-0.15}^{+0.16}$ & 4.09$\pm 0.16$ & 0.66$\pm 0.06$ & 3248.37 \\
         & 0.33$_{-0.10}^{+0.11}$ & 5.0$_{-0.6}^{+0.7}$ & 16.8$\pm 6.9$ & 0.03$\pm 0.02$ & 3237.69 \\
\\
SHspot & -- & 2.16$_{-0.08}^{+0.08}$ & 7.2$\pm 0.2$ & 6.5$\pm 0.9$ & 3849.86  \\
         & 0.06$_{-0.04}^{+0.04}$ & 2.24$_{-0.15}^{+0.16}$ & 7.7$\pm 1.3$ & 8.6$\pm 1.1$ & 3849.59 \\
\\
SJet & -- & 2.2$_{-0.2}^{+0.2}$ & 1.42$\pm 0.07$ & 1.24$\pm 0.62$ & 3279.82  \\
         & $<0.08$ & $<2.1$  & 3.3$\pm 2.0$ & 0.68$\pm 0.47$ & 3279.47 \\
\\
SLobe & -- & 2.73$_{-0.23}^{+0.24}$ & 2.5$\pm 0.1$ & 0.97$\pm 0.50$ & 3885.83  \\
         & $<0.13$ & $<2.8$ & 3.6$\pm 1.9$ & 0.60$\pm 0.3$ & 3885.77 \\
\\
SLobe1 & -- & 2.1$_{-0.3}^{+0.3}$ & 1.18$\pm 0.06$ & 1.18$\pm 0.66$ & 3422.7 \\
         & $<0.046$ & $<2.0$  & 18.6$\pm 9.7$ & $<0.4$ & 3421.85 \\
\\
Center & -- & 1.15$_{-0.06}^{+0.06}$ & 9.5$\pm 0.2$ & 39.9$\pm4.5$ & 5377.53  \\
         & $<0.001$ & $<1.1$  & 9.1$\pm 0.6$ & 37.1$\pm 4.5$ & 5320.22 \\
\\
EArm & -- & 3.5$_{-0.3}^{+0.3}$ & 1.58$\pm 0.14$ & 0.2$\pm0.07$ & 3198.87 \\
  & 0.65$_{-0.30}^{+0.34}$ & 7.7$_{-2.1}^{+2.3}$ & 50.1$\pm5.2$ & 0.02$\pm 0.01$ & 3193.89 \\
\\
HBranch & -- & 3.5$_{-0.2}^{+0.2}$ & 2.6$\pm 0.2$ & 0.45$\pm 0.01$ & 3856.6  \\
         & 0.60$_{-0.18}^{+0.22}$ & 7.6$_{-1.4}^{+1.5}$ & 98.0$\pm39.8$ & 0.04$\pm0.03$ & 3845.09\\
\\
HBranchS & -- & 3.0$_{-0.2}^{+0.2}$ & 4.21$\pm 0.47 $ & 1.09$\pm 0.21$ & 3838.9 \\
         & 0.53$_{-0.16}^{+0.17}$ & 6.3$_{-1.1}^{+1.2}$ & 58.2 $\pm 17.0$ & 0.15$\pm0.09$ & 3829.35 \\
\\
\\
Core\_ext & -- & 2.5$_{-0.2}^{+0.2}$ & 7.95$\pm 0.18$ & 4.1$\pm 1.5$ & 4422.43 \\
& $<$0.002 & $<2.2$ & 6.83$\pm0.42$ & 14.6$\pm2.8$ & 4407.48 \\
\\
\hline
\end{tabular}
\end{center}
NOTE.- The first row for each regions shows the best fit model parameters for the absorption fixed
at the Galactic column of 3.98$\times 10^{20}$\,cm$^{-2}$. The second row shows the best fit
absorption column for that region. Errors are 68\% for one significant parameter and upper and lower limits are at 90\%\\
$^a$ Hydrogen equivalent absorption column in 10$^{22}$\,cm$^{-2}$. It was frozen in the first fit and  thawed in the second fit for each region;\\
$^b$ Model flux in the units of 10$^{-15}$\,erg\,cm$^{-2}$\,s$^{-1}$: Errors calculated via simulations with {\tt {sample\_energy\_flux}} in Sherpa.\\
$^c$ degrees of freedom were equal to 3565 and 3564 for the model fit with a frozen and thawed absorption parameter
respectively.

}
\end{table*}


%% file: table5.tex
\begin{table*}
{\scriptsize
\noindent
\caption[]{\label{table-5} Parameters of Thermal Model Fits to X-ray Spectra of Various Regions.}
\begin{center}
\begin{tabular}{ccccccccccccccc}
\hline\hline
\\
Region & N$_H$$^a$ & kT$^b$ & Abundance$^c$ &  Norm$^d$ & Flux(0.5-2)$^e$ &  Flux(2-10))$^e$ & Stat (Cash)$^f$ \\
\\
\hline
\\
NLobe & -  & 0.45 $^{+ 0.05 }_{-0.04 }$ &  - & 10.84 $^{+ 2.19}_{-1.76}$ &  4.47$\pm 0.46$ & 0.15$\pm 0.03$ & 3519.1 & \\
      & $0.157^{+0.090}_{-0.008}$  &  0.33 $^{+ 0.09 }_{-0.08 }$ & - & 33.19$^{+0.08}_{-0.06}$ & 7.02$\pm 0.22$ & 0.05$\pm 0.01$ & 3516.8 & \\
      & - & 0.64 $^{+ 0.10 }_{ -0.06 }$ & 0.06 $^{+ 0.03}_{ -0.02}$ & 13.4 $^{+ 3.7 }_{-3.6 }$ & 4.2$\pm 0.4$ & 0.20$\pm 0.05$ & 3516.8 \\
NLobe1 & -   & 1.9 $^{+ 1.0 }_{-0.5 }$ & - & 0.55$^{+ 0.14 }_{-0.11 }$ &  0.81$\pm0.12$  & 0.44$\pm0.15$ & 3081.8 & \\
         & $ <0.1 $ & $2.1^{+1.1}_{-0.7}$ & - & $0.18^{+0.08}_{-0.05}$ & 0.72$\pm 0.19$ & 0.48$\pm0.15$ & 3081.7 & \\
         & - &  2.0 $^{+ 1.2 }_{ -0.6 }$ & $<0.8$ & 1.8 $^{+ 0.5 }_{-0.4 }$ &  0.8$\pm0.1$ & 0.4$\pm0.3$ & 3086.9 \\
NHspot & -   & 0.54 $^{+ 0.06 }_{-0.04 }$ & - & 7.28 $^{+ 1.17 }_{-1.15 }$ & 3.97$\pm 0.28$ & 0.19$\pm0.03$ & 3240.9 & \\
         & 0.078 $^{+ 0.067 }_{-0.058 }$ & 0.49$^{+ 0.07 }_{-0.06 }$ & - & 9.75 $^{+1.07 }_{-0.08 }$ & 4.64$\pm 0.20$ & 0.16$\pm0.03$ & 3240.5 & \\
       & - & 0.79 $^{+ 0.06 }_{ -0.06 }$ & 0.06 $^{+ 0.03}_{ -0.02}$ & 9.4 $^{+ 2.3 }_{-1.7 }$ & 3.7$\pm0.3$ & 0.29$\pm0.07$  & 3233.9 \\
SHspot & -   & 2.2 $^{+ 0.3 }_{-0.2 }$ & - & 4.58 $^{+ 0.31 }_{-0.28 }$ & 7.05$\pm 0.38$ & 4.65$\pm0.10$ & 3855.7 & \\
         & $< 0.01$ & $<2.57 $ & - & 3.95$^{+ 0.24 }_{-0.28 }$ & 6.29$\pm 0.38$ & 4.96$\pm 0.10$ & 3850.0 & \\
       & - & 2.3 $^{+ 0.3 }_{ -0.2 }$ & $<0.09$ & 15.8 $^{+ 0.9 }_{-1.0 }$ & 7.0$\pm 0.5$& 4.7$\pm0.1$ &  3860.2 \\
SJet & -   & 2.1 $^{+ 0.9 }_{-0.6 }$ & - & 0.91$^{+ 0.18 }_{-0.14 }$ & 1.37$\pm 0.07$ & 0.84$\pm0.17$ & 3283.9 & \\
         & $<0.022$  & $<2.6 $ & - & $<1.7$ & 1.22$\pm 0.07$ & 0.96$\pm 0.19$ & 3282.0 & \\
     & - & 2.3 $^{+ 0.9 }_{ -0.6 }$ & $<0.13$ & 3.1 $^{+ 0.6 }_{-0.4 }$ & 1.4$\pm0.2$& 0.9$\pm0.7$& 3289.2 \\
SLobe & -   & 0.80 $^{+ 0.21 }_{-0.15 }$  & - &  2.82 $^{+ 0.78 }_{-0.59 }$ & 2.41$\pm 0.13$ & 0.31$\pm0.08$ & 3887.4 & \\
         & $< 0.042$ & $<1.1 $ & - & 2.08 $^{+ 0.27 }_{-0.22 }$ & 2.09$\pm 0.13$ & 0.39$\pm 0.08$ & 3886.4 & \\
      & - & 0.9 $^{+ 0.2 }_{ -0.2 }$ &  $<0.03$ & 8.9 $^{+ 2.3 }_{-1.7}$ & 2.4$\pm0.3$& 0.4$\pm0.2$& 3893.9 \\
SLobe1 & -   & 2.24 $^{+ 1.40 }_{-0.74 }$ & - & 0.74 $^{+ 0.18 }_{-0.13}$ & 1.15$\pm 0.06$ & 0.77$\pm 0.16$ & 3424.6 & \\
         & $<0.027$ & $<2.6$  &  & $<2.4$ & 1.05$\pm 0.06$ & 0.82$\pm 0.16$ & 3423.7 & \\
      & - & 2.45 $^{+ 1.98 }_{ -0.75 }$ & $<1.5$ & 2.6 $^{+ 0.5 }_{-0.7 }$ & 1.2$\pm0.2$& 0.81$\pm0.08$ & 3430.6 \\
Center$^g$ & -   & 0.53$^{+0.13}_{-0.02}$ & - & 19.9$^{+13.9}_{-7.6}$ & 9.7$\pm 0.5$  &  35.7$\pm 7.1$ & 4685.7 & \\
        & $<19.71$ & 0.63$^{+0.08}_{-0.24}$ & - & 15.7$^{+17.3}_{-3.3}$ & 10.4$\pm 0.6$ & 33.5$\pm 6.8$ & 4684.5 & \\
       & - & 18.7 $^{+ 4.8 }_{ -2.6 }$ & $>4.6$  & 15.1 $^{+ 0.9 }_{-0.7 }$ &  10.2$\pm1.1$& 33.1$\pm3.1$& 5379.2 \\
Total & -   & 0.59 $^{+ 0.03 }_{-0.03 }$ & - & 51.38 $^{+ 4.17 }_{-3.82 }$ & 31.2$\pm 1.6$ & 1.9$\pm 0.1$ & 1491.9 & \\
         &  $<0.006$ & $<0.74$  & - & $<34.24 $   & 26.8$\pm 1.6$ & 2.8$\pm 0.1$ & 1485.5 & \\
      & - & 0.78 $^{+ 0.03 }_{ -0.03 }$ & 0.04 $^{+ 0.01 }_{ -0.01}$ & 91.5 $^{+ 6.4 }_{-5.9 }$ & 29.8$\pm1.1$ & 2.6$\pm0.3$ & 1492.4 \\
EArm & -   & 0.39 $^{+ 0.08 }_{-0.06 }$ & - & 4.74$^{+ 2.01 }_{-1.40}$ & 1.54$\pm 0.78$ & 0.02$\pm 0.01$ & 3195.6 & \\
         & 0.44 $^{+ 0.28 }_{-0.20 }$ & $<0.27$ & - & $<314.9$  & 12.8$\pm 0.1$ & $<0.001$ & 3190.8 & \\
     & - & 0.71 $^{+ 0.08 }_{ -0.16 }$ & 0.13 $^{+ 0.14 }_{ -0.08 }$ & 2.5 $^{+ 2.1 }_{-1.0 }$ & 1.3$\pm0.1$ & 0.07$\pm0.01$ & 3192.7 \\
HBranch & -   & 0.40 $^{+ 0.05 }_{-0.04 }$ & - & 10.63$^{+ 2.84 }_{-2.19 }$ & 3.52$\pm 0.16$ & 0.05$\pm 0.01$ & 3845.3 & \\
         & 0.573 $^{+ 0.242 }_{-0.177 }$ & $<0.24$   & - & $<305.0$  & 63.8$\pm 3.8 $ & $<0.001$ & 3832.5 & \\
      & - & 0.75 $^{+ 0.05 }_{ -0.06 }$ & 0.18 $^{+ 0.13}_{ -0.07}$ & 4.7 $^{+ 1.8 }_{-1.6 }$ & 3.0$\pm0.4$ & 0.15$\pm0.06$ & 3832.9 \\
HBranchS & -   & 0.51 $^{+ 0.06 }_{-0.05 }$ & - &  8.19 $^{+ 1.66}_{-1.28 }$ & 4.15$\pm 0.21$ & 0.16$\pm 0.03$ & 3834.3 & \\
         & 0.327 $^{+ 0.114 }_{-0.102 }$ & 0.27 $^{+ 0.11 }_{-0.10 }$ & - & $<121.0$  & 15.7$\pm 1.6$ & 0.03$\pm 0.01$ & 3824.3 & \\
      & - & 0.75 $^{+ 0.06 }_{ -0.06 }$ & 0.09 $^{+ 0.04 }_{ -0.03 }$ & 8.4 $^{+ 2.1 }_{-1.8 }$ & 3.8$\pm0.4$ & 0.25$\pm0.06$ & 3817.3 \\
Core$\_{\rm ext}$ & -   & 8.2 $^{+ 10.2 }_{-2.7 }$ & - & 3.98 $^{+ 0.47 }_{-0.18 }$ & 6.6$\pm 0.4$  & 13.1$\pm 2.5$ & 4515.1 & \\
         & $<0.002$ & $<$12.05  & - & $<15.8$ & 5.9$\pm 0.4$ & 14.1$\pm 2.8$ & 4472.9 & \\
         & - & 13.7 $^{+ 3.2 }_{ -1.3}$ & $>3.7$ & 9.8 $^{+ 0.8 }_{-0.5 }$ & 7.2$\pm0.2$ & 22.3$\pm2.2$ & 4766.5 \\
\\
\hline
\end{tabular}
\end{center}
NOTE.- The first row for each regions shows the best fit model parameters for the absorption fixed at the Galactic column of 3.98$\times 10^{20}$~\,cm$^{-2}$ and the abundance fixed at 0. The second row shows the best fit absorption column for that region. The third row shows the results of the model fit with abundance free
and the absorption column fixed at the Galactic value. Errors are 68\% for one significant parameter and upper and lower limits are at 90\%.\\
$^a$ Hydrogen equivalent absorption column in 10$^{22}$\,cm$^{-2}$; 
$^b$ the temperature in keV;
$^c$ metal abundances in terms of fraction of the Solar;
$^d$ Norm in units of 10$^{-6}$\,photons\,cm$^{-2}$\,s$^{-1}$; 
$^e$ Flux in the units of 10$^{-15}$\,erg\,cm$^{-2}$\,s$^{-1}$; Errors calculated via simulations.
$^f$ degrees of freedom were equal to 3565 and 3564 for the model fit with a frozen and thawed absorption/abundance parameter
respectively; 
$^g$ Center - an additional power law model component for this region, with the best fit parameters $\Gamma= -3.47^{+1.79}_{-0.07}$ and $\Gamma =-0.29^{+0.18}_{-3.21}$ for fixed and thaw absorption respectively.

}
\end{table*}

%% file: table6.tex
\begin{table*}
{\scriptsize
\noindent
\caption[]{\label{table-6} Physical Parameters of the Regions$^a$}
\begin{center}
\begin{tabular}{ccccccccccccccc}
\hline\hline
\\
Region & Norm$^b$ & Area$^c$ &  $n_e^d$ &  $P_{th}^e$ \\
\\
& [10$^{-6}$ photons~cm$^{-2}$~s$^{-1}$] & [arcsec$^2$] & [cm$^{-3}$] & [10$^{-10}$ dyne~cm$^{-2}$] \\
\\
\hline
\\
NLobe & 10.84$\pm2.19$ & 20.82 & 0.18$\pm0.02$ & 2.60$\pm0.39$ \\
\\
NLobe1 & 0.552$\pm0.145$ & 23.93 & 0.038$\pm0.005$ & 2.31$\pm1.25$ \\
\\
NHspot & 7.28$\pm1.17$ & 8.08 & 0.24$\pm0.02$ & 4.11$\pm0.56$ \\
\\
SHspot & 4.58$\pm0.31$ & 9.55 & 0.17$\pm0.01$ & 12.20$\pm1.5$ \\
\\
SJet & 0.91$\pm0.18$ & 30.69 & 0.043$\pm0.004$ & 2.90$\pm1.27$ \\
\\
SLobe & 2.82$\pm0.78$ & 251.13 & 0.026$\pm0.004$ & 0.68$\pm0.19$ \\
\\
SLobe1 & 0.74$\pm0.18$& 91.02 & 0.022$\pm0.003$ & 1.62$\pm1.03$ \\
\\
Center & 19.9$\pm13.9$ & 75.93 & 0.128$\pm0.045$ & 2.17$\pm0.93$ \\
\\
Total & 51.38$\pm4.16$ & 1751.77 & 0.043$\pm0.002$ & 0.81$\pm0.05$ \\
\\
EArm & 4.74$\pm2.01$ & 58.73 & 0.071$\pm0.015$ & 0.91$\pm0.26$ \\
\\
HBranch & 10.63$\pm2.84$ & 192.99 & 0.059$\pm0.008$ & 0.75$\pm0.14$ \\
\\
HBranchS & 8.19$\pm1.66$ & 115.44 & 0.07$\pm0.01$ & 1.07$\pm0.17$ \\
\\
Core & 4.2$\pm0.5$ & 4.91 & 0.23$\pm0.01$ & 50.41$\pm18.77$ \\
\\
Core\_ext & 3.98$\pm0.47$ & 57.45 & 0.066$\pm0.004$ & 17.29$\pm10.59$ \\
\\
\hline
\end{tabular}
\end{center}

$^a$ Density and thermal pressure in the selected regions calculated using the results of a 
thermal model fit to the spectra. 

$^b$ Normalization of thermal bremsstrahlung model in units of 10$^{-6}$ photons~cm$^{-2}$~s$^{-1}$.
Norm$=\frac{3.02\times10^{-15}}{4\pi\ D^2} \int n_e n_I dV$, where $n_e$
is the electron density (cm$^{-3}$), $n_I$ is the ion density
(cm$^{-3}$), and $D$ is the distance to the source (cm);

$^c$ Area in units of arcsec$^2$. 
The third dimension assumed in calculation of the electron density was set to 1\,kpc.

$^d$ Electron density in cm$^{-3}$. 1$\sigma$ errors due to the uncertainty on normalization.

$^e$ Thermal pressure in units of 10$^{-10}$\,dyne~cm$^{-2}$. 1$\sigma$ errors due to the normalization  and temperature errors only.

}
\end{table*}